\RequirePackage[2020-02-02]{latexrelease}
\documentclass[12pt, notitlepage,aps,preprintnumbers,superscriptaddress,nofootinbib,aps,prd,floatfix]{revtex4-2}
\usepackage{float}
\usepackage{physics,slashed}
\usepackage{amsfonts,amsmath,amssymb}
\usepackage{mathrsfs}
\usepackage{bm}

\usepackage{epsfig}
\usepackage{graphicx}   
\usepackage{subfigure}  
\usepackage{verbatim}   
\usepackage{color}      
\usepackage{hyperref}   
\usepackage{url}
\definecolor{nicered}{rgb}{0.5,0.,0.}
\definecolor{nicegreen}{rgb}{0.,0.5,0.}
\definecolor{niceblue}{rgb}{0.,0.,0.5}
\hypersetup{colorlinks,citecolor=nicegreen,linkcolor=nicered,urlcolor=niceblue}
\usepackage{array}
\usepackage{dcolumn}
\usepackage{multirow}
\usepackage{cprotect}
\usepackage{framed}
\usepackage{setspace}

\newcommand{\MeV}{\textrm{MeV}}
\newcommand{\GeV}{\textrm{GeV}}

\begin{document}
	\title{Understanding PDF uncertainty on the $W$ boson mass measurements in CT18 global analysis}
	\author{Jun Gao}
	\email{jung49@sjtu.edu.cn}
	\affiliation{
		NPAC, Shanghai Key Laboratory for Particle Physics and Cosmology, \\
		School of Physics and Astronomy, Shanghai Jiao Tong University, Shanghai 200240, China}
	\affiliation{
		Key Laboratory for Particle Astrophysics and Cosmology (MOE), Shanghai 200240, China}
	\author{DianYu Liu}
	\email{dianyu.liu@sjtu.edu.cn}
	\affiliation{
		NPAC, Shanghai Key Laboratory for Particle Physics and Cosmology, \\
		School of Physics and Astronomy, Shanghai Jiao Tong University, Shanghai 200240, China}
	\affiliation{
		Key Laboratory for Particle Astrophysics and Cosmology (MOE), Shanghai 200240, China}
	\author{Keping Xie}
	\email{xiekeping@pitt.edu}
	\affiliation{Pittsburgh Particle Physics, Astrophysics, and Cosmology Center,
		Department of Physics and Astronomy, University of Pittsburgh, Pittsburgh, PA 15260, USA\looseness=-1}
	
	\preprint{PITT-PACC-2209}
	
	\date{\today}
	\begin{abstract}
We study the dependence of the transverse mass distribution of the charged lepton and the missing energies on the parton distributions (PDFs) adapted to the $W$ boson mass measurements at the CDF and ATLAS experiments. 
We compare the shape variations of the distribution induced by different PDFs and find that spread of predictions from different PDF sets can be much larger than the PDF uncertainty predicted by a specific PDF set. 
We suggest analyzing the experimental data using up-to-date PDFs for a better understanding of the PDF uncertainties in the $W$ boson mass measurements. 
We further carry out a series of Lagrange multiplier scans to identify the constraints on the transverse mass distribution imposed by individual data sets in the CT18 global analysis. 
In the case of CDF measurement, the distribution is mostly sensitive to the $d$-quark PDFs at the intermediate $x$ region that is largely constrained by the DIS and Drell-Yan data on the deuteron target, as well as the Tevatron lepton charge asymmetry data.		
	\end{abstract}
	
	\maketitle
	\tableofcontents
	\section{Introduction}
	Recently, the CDF collaboration released their latest high-precision measurement
	of the $W$ boson mass with the Tevatron 1.96 TeV proton-antiproton collision with
	8.8 fb$^{-1}$ integrated luminosity~\cite{CDF:2022hxs}. 
	The reported result is
	\begin{equation}
		M_W=80,433.5\pm6.4_{\rm stat}\pm6.9_{\rm syst}~\MeV,
	\end{equation}
	which departs from the Standard Model electroweak precision fit $M_W=80,357\pm6~\MeV$\footnote{The HEPfit group gives a consistent global fit as $M_W=80,354.5\pm5.7~\MeV$~\cite{deBlas:2021wap}.}~\cite{Haller:2018nnx,ParticleDataGroup:2020ssz}
	by 7 standard deviations. 
	This new measurement triggers a lot of discussions concerning possible new physics effects (see Refs.~\cite{2204.03796,2204.03767,2204.03693,2204.04191,2204.04202,2204.04356,2204.04286,Campagnari:2022vzx,2204.04514,2204.04559,2204.04770,2204.04688,2204.04672,2204.04834,2204.05267,2204.05303,2204.05296,2204.05284,2204.05302,2204.05024,2204.05031,2204.05269,2204.05085,2204.05283,2204.05285,2204.05728,2204.05962,2204.05965,2204.05760,2204.05942,2204.05945,2204.05975,2204.06327,2204.06541,2204.06485,2204.06234,2204.07022,2204.07100,2204.07138,2204.07144,2204.07411,2204.07511,2204.07844,2204.07970,2204.08067,2204.08440,2204.08266,2204.08390,2204.08406,2204.08568,2204.08546,2204.09031,2204.09001,2204.09024,2204.09487,2204.09376,2204.09671,2204.09477,2204.09585,2204.10274,2204.10315,2204.10156,2204.10130,2204.10338,2204.11570,2204.11755,2204.11871,2204.11945,2204.11974,2204.11991,2204.12018,2204.12152,2204.12898,2204.13272,2204.13690,2205.00758,2205.00783,2205.01115,2205.01437,2205.01701,2205.01699, 2205.02088,2205.02205,2205.02217,2205.01911} as examples), Standard Model Effective Theory (SMEFT) implications~\cite{2204.04204,2204.04805,2204.05260,2204.05992}, as well as evaluations of various theoretical uncertainties~\cite{2204.03996,2204.12394,2204.13500,2205.00031,2205.02788}.
	Previously, there are also direct measurements at the LHC 7 TeV from the ATLAS
	collaboration with $M_W=80,370\pm7_{\rm stat}\pm18_{\rm syst}$~\MeV~\cite{ATLAS:2017rzl},
	at the LHC 13 TeV from the LHCb collaboration with
	$M_W=80,354\pm23_{\rm stat}\pm22_{\rm syst}$~\MeV~\cite{LHCb:2021bjt},
	and ealier from Tevatron~\cite{CDF:2013dpa} and LEP~\cite{ALEPH:2013dgf}.
	While there are discussions concerning disagreements between the new CDF measurement and
	previous ones, in this study we only focus on the impact of the parton distributions (PDFs)
	on the extracted $W$ boson mass.
	In the CDF, ATLAS, and LHCb measurements mentioned earlier, they reported PDF
	uncertainties of about 3.9, 8, and 9 MeV respectively, being one of the dominant
	theoretical uncertainties in all cases. 
	We note the three collaborations use very different inputs for the evaluation of the PDF
	uncertainties.
	For example, the quoted PDF uncertainty from CDF is purely based on NNPDF3.1 PDFs~\cite{NNPDF:2017mvq}
	though the impact of different PDFs has been investigated.
	For the ATLAS measurement, the quoted PDF uncertainty includes those estimated from
	CT10 PDFs~\cite{Gao:2013xoa} adding in quadrature with differences observed for using PDFs from several groups. 
	In the LHCb measurement, they take the arithmetic average of the PDF uncertainties
	predicted by CT18~\cite{Hou:2019efy}, NNPDF3.1~\cite{NNPDF:2017mvq} and MSHT20~\cite{Bailey:2020ooq} PDFs.
	LHCb also reports a spread of central values of the extracted $W$ boson mass from the three
	PDFs, which is as large as 11 MeV and is not counted in the final uncertainty.
	The way that PDFs change the modeling on the kinematics of the decayed leptons in the $W$
	boson production can be understood as below.
	The fully differential cross sections of the decayed leptons can be written as
	\begin{align}~\label{eq:fac}
		\frac{\dd^2\sigma}{\dd p_1\dd p_2}=&\left[\frac{\dd\sigma(m)}{\dd m}\right]\left[\frac{\dd\sigma(y)}{\dd y}\right]
		\left[\frac{\dd^2\sigma(p_T,y)}{\dd p_T\dd y}\left(\frac{\dd\sigma(y)}{\dd y}\right)^{-1}\right]\nonumber\\
		& \times \left[(1+\cos^2\theta)
		+\sum_{i=0}^7 A_i(p_T,y)P_i(\cos\theta,\phi)\right],
	\end{align}
	where $p_{1(2)}$ are the lepton and anti-lepton momentum.
	$m$, $p_T$, and $y$
	are the invariant mass, transverse momentum, and rapidity of the dilepton system.
	$\theta$ and $\phi$ are the polar angle and azimuth of $p_1$ in the rest frame of the dilepton system.
	$A_i$ are angular coefficients, and $P_i$ are spherical harmonics.
	The cross sections are factorized in such a way since in the experimental analyses
	each part on the right-handed side of Eq.~(\ref{eq:fac}) is modelled or corrected
	separately by using different MC programs~\cite{ATLAS:2017rzl}. 
	The impact of PDF uncertainties on the individual components can be understood. 
	For example, the effects of PDF variations on the angular coefficients and the
	invariant mass distribution is found to be small in the $W$ boson mass
	measurements.
	Most of the PDF uncertainties on the extracted $W$ boson mass
	arise due to the impact on the rapidity distribution of the $W$ boson.
	The PDF variations especially those from the gluon PDF will also affect the
	transverse momentum distribution at large-$p_T$.
	Complication about the $p_T$ distribution is that the experimental analyses use
	a data-driven method by normalizing it to their measured transverse momentum distribution
	of the $Z$ boson, in order to reduce the theoretical uncertainties.
	Thus concerning the component of $p_T$ distribution, only the theoretical uncertainties
	on the ratio of the $p_T$ spectrum of the $W$ boson and the $Z$ boson should be included. 
	In this study, we carry out a comprehensive comparison of the kinematic distribution
	of the decayed leptons using a variety of up-to-date PDFs focusing on the CDF and the
	ATLAS measurement.
	There also exist previous studies on the PDF uncertainties in the $W$ boson mass
	measurements~\cite{Nadolsky:2004vt,Bozzi:2015hha,Farry:2019rfg,1910.04726,Hussein:2019kqx}. 
	We show variations in the shape of the transverse mass distribution from
	different PDF groups as well as from different generations within the same
	group, and estimate the possible shift on the extracted $W$ boson mass.
	Furthermore, using the method of Lagrange multiplier (LM) scan, we evaluate the
	constraints on the prescribed distribution as imposed by different experimental data sets
	in the CT18 global analysis of PDFs.
	Other theoretical uncertainties, including factorization and renormalization scales, 
	the strong coupling constant, and the $W$-boson decay width are examined as well in the CDF scenario.
	The rest of our paper is organized as follows. 
	In Sec.~\ref{sec:A}, we present results on the kinematic
	distribution and on the extracted $W$ boson mass.
	In Sec.~\ref{sec:B}, we show results on PDF sensitivities and
	on the understanding of constraints within CT18 analysis using the LM method.
	Finally, our summary and conclusions are presented in Sec.~\ref{sec:C}. 

	\section{PDF dependence of the kinematic distribution}~\label{sec:A}
	In this section, we show the dependence of the transverse mass distribution of the
	charged lepton and the missing energies on the PDFs, as well as on
	the change of the $W$ boson mass, for both the setups of the CDF measurement
	and the ATLAS 7 TeV measurement.
	We note that in both the CDF and ATLAS measurements the PDF uncertainties are
	fully correlated for results using three different kinematic variables,
	and are almost the same in size.
	We calculate the mean value of the transverse mass to
	quantify the impact of different PDFs on the shape of the kinematic distribution.
	We further propose a simplified prescription to identify PDF impact on the
	extracted $W$ boson mass and validate it against a method of using log-likelihood
	$\chi^2$ fit.
	We clarify that the purpose of our study is trying to understand
	the PDF uncertainties in the $W$ boson mass measurements especially
	using the most up-to-date PDFs, rather than to
	reproduce exact PDF dependence in the actual experimental analyses. 

	\subsection{The CDF measurement}\label{sec:cdf}
	The event selection criterion follows the CDF Run II measurement~\cite{CDF:2022hxs}
	\begin{equation}
		30<p_T^{\ell,\nu}<55~\GeV,~u_T<15~\GeV,~ 60<M_T<100~\GeV,
	\end{equation}
	where $u_T=|\vec{p}_T^\ell+\vec{p}_T^\nu|$ is the transverse momentum of $W$ boson.
	The transverse mass of $W$ boson is defined $M_T=\sqrt{2(p_T^{\ell}p_T^{\nu}-\vec{p}_T^{\ell}\cdot\vec{p}_T^{\nu})}$.
	The charged lepton is further required to be in the central pseudo-rapidity
	region
	\begin{equation}
		|\eta|<1.
	\end{equation}
	The transverse mass distribution of the charged lepton and missing energies are
	calculated using the program MCFM-6.8~\cite{Campbell:1999ah,Campbell:2011bn} with above selections and at either
	leading order (LO) or next-to-leading order (NLO) in QCD, and with APPLgrid
	interface~\cite{Carli:2010rw} for fast interpolations with arbitrary PDFs. 
	We have not considered the QCD resummation effects on the transverse momentum of the
	$W$ boson as they are supposed to be less pronounced for the $M_T$ distribution and
	also because constraints of repeating the calculations for a large number of PDF sets
	with sufficient numerical accuracy.
	For the same reasons we have not included the next-to-next-to-leading order (NNLO) QCD
	corrections and we expect they will not change our conclusion concerning PDF dependence
	much. 
	The factorization and renormalization scales are chosen to be invariant mass 
	of charged lepton and neutrino pair, $\mu_F=\mu_R=M_{\ell\nu}$. 
	We apply a Gaussian smearing on distributions from
	theoretical calculations assuming a detector resolution of 7\% for $M_T$,
	which is consistent with the resolution on hadronic recoils reported in the CDF paper~\cite{CDF:2022hxs}.
	We check that the prescribed smearing effects can reproduce well the shape
	of the experimental distribution especially for close to the peak region.
	In principle, one can apply more sophisticated detector effects with energy-dependent
	resolution for individual objects.
	We do not expect a large difference especially since we are focusing
	on the mean value of the kinematic distribution that is less affected by smearing
	effects.
	\begin{figure}
		\includegraphics[width=0.49\textwidth]{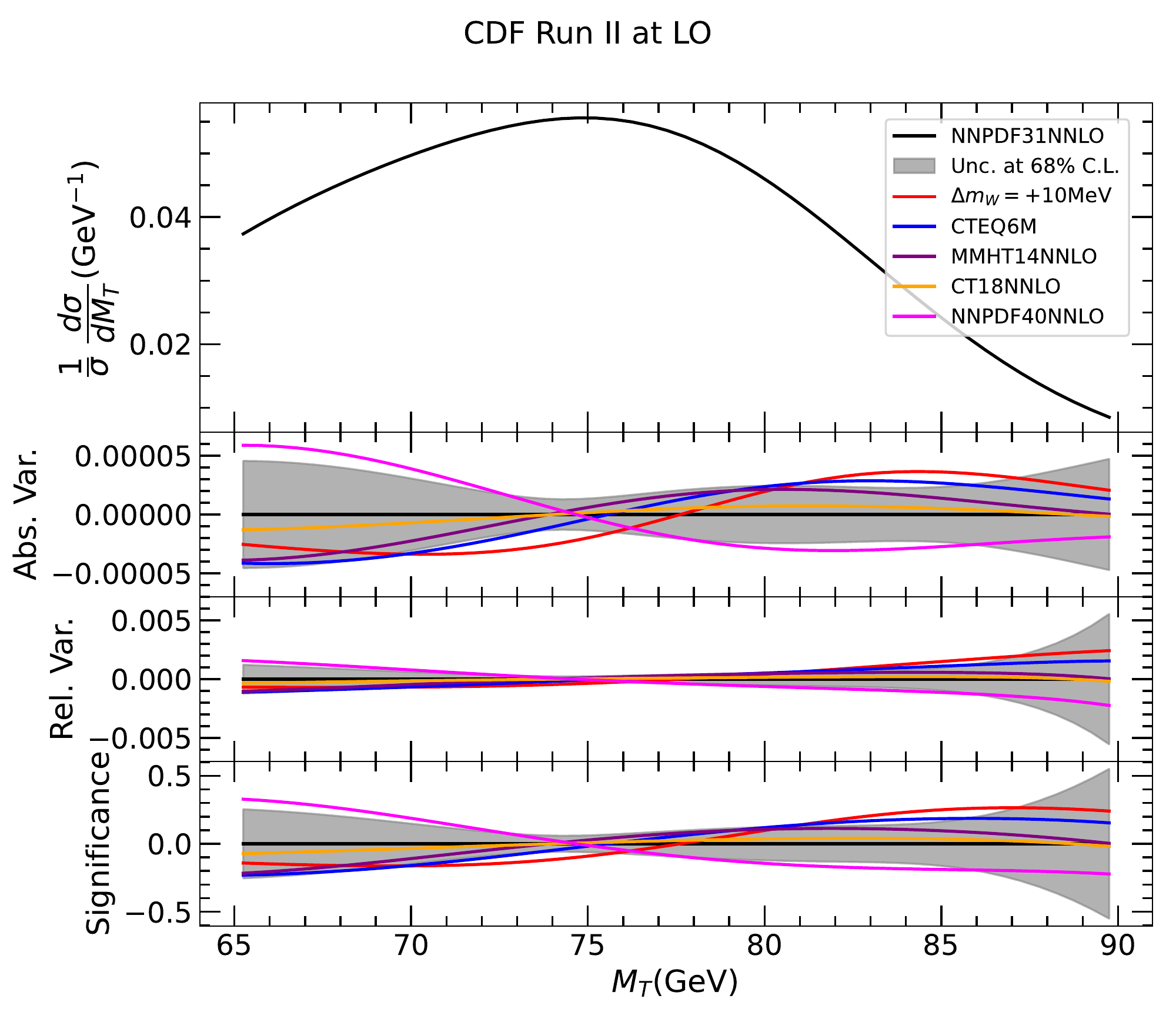}
		\includegraphics[width=0.49\textwidth]{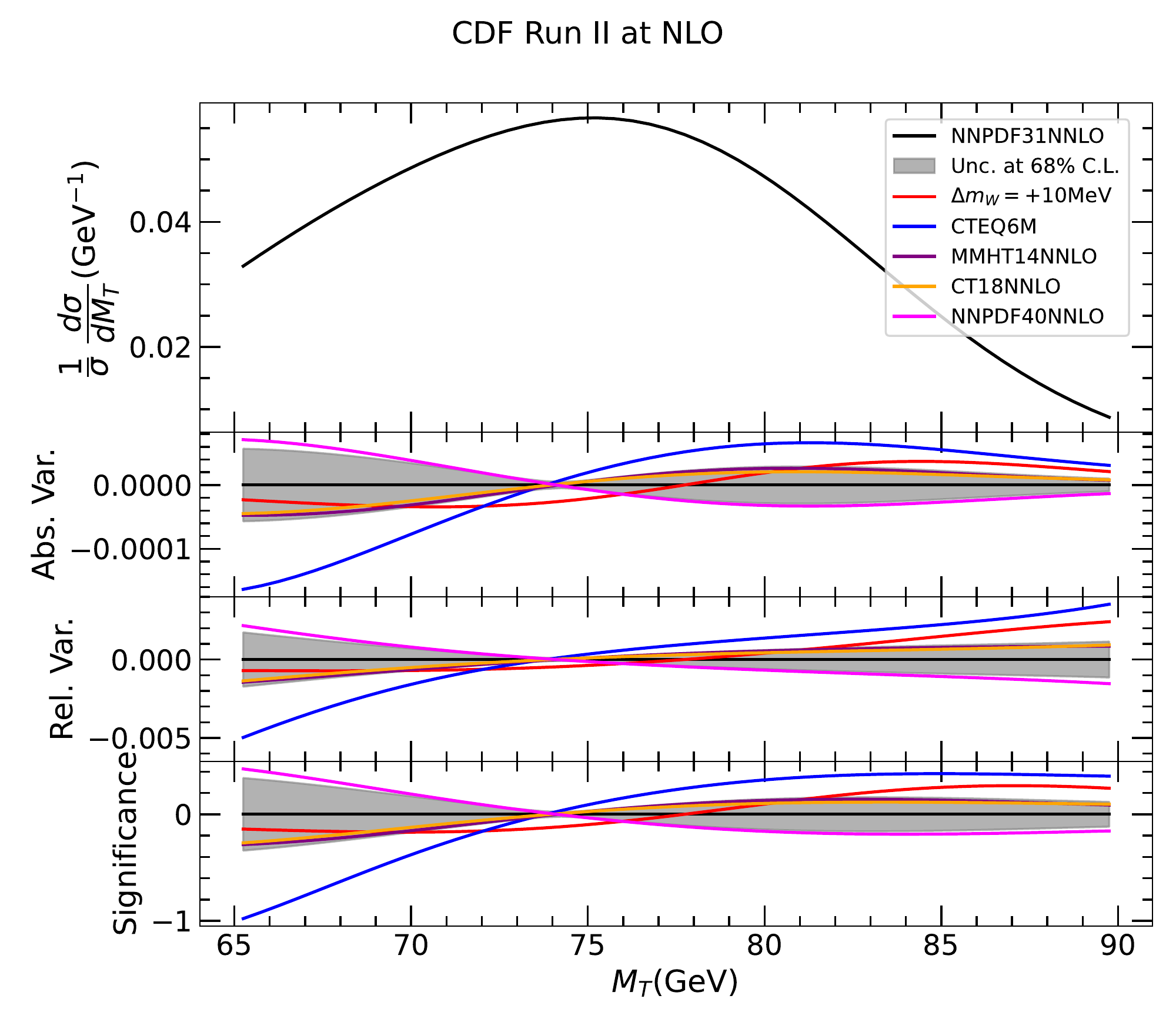}
		\caption{
			Transverse mass distribution of the charged lepton and missing energies
			for the scenario of CDF measurement calculated at LO and NLO with various PDFs
			and different choices of the $W$ boson mass (increased by 10 MeV). 
			From top to bottom 
			are the normalized distribution, absolute and relative changes with respect
			to a common reference of the prediction
			obtained with NNPDF3.1 NNLO PDFs and nominal $W$ boson mass.
			The lowest panel shows the changes normalized to the experimental
			statistical uncertainties.
		}
		\label{fig:cdfdis}
	\end{figure}

	In Fig.~\ref{fig:cdfdis} we show the predictions on the normalized $M_T$ distribution
	at LO and NLO for several choices of the PDFs and with different choices of the $W$
	boson mass.
	From top to bottom it shows the normalized distribution, the absolute and relative
	variations compared to a common reference calculated with the central set of
	NNPDF3.1 NNLO PDFs~\cite{NNPDF:2017mvq}.
	The red curve represents the variation due to a $W$ boson mass change of +10 MeV,
	and the gray band indicates the PDF uncertainties at 68\% Confidence Level (C.L.) for NNPDF3.1.
	Different PDFs considered include CTEQ6M~\cite{Stump:2003yu} NLO PDFs, and CT18~\cite{Hou:2019efy}, 
	MMHT14~\cite{Harland-Lang:2014zoa} and NNPDF4.0~\cite{Ball:2021leu} NNLO PDFs.
	In the lower panel of each figure, the variations are divided by the statistical
	uncertainty in each bin to show the significance.
	We choose a bin width of 0.5 GeV and normalize the total number of events to the
	number of muon events in the CDF measurement for the calculation of the statistical
	uncertainty.
	There are several interesting observations.
	First, we found there can be a significant difference in the PDF dependence when
	going from LO to NLO, while the variation due to $W$ boson mass is much more stable.
	That can be understood as due to the gluon contributions at NLO which boost
	the $W$ boson in the transverse direction.
	Especially, the gluon PDF is quite different in CTEQ6M compared to the recent NNLO
	PDFs which leads to the large differences seen in the NLO plot.
	That can be traced back to the fact that the CTEQ6M analysis uses a zero-mass
	scheme for the heavy-quark effects in DIS rather than variable flavor number schemes.
	However, we stress that in the experimental analyses since they use data on
	$Z$ boson $p_T$ spectrum to model the $W$ boson $p_T$ spectrum, only PDF uncertainties
	on the ratio of the $W$ and $Z$ boson $p_T$ spectrum should be considered concerning the
	$p_T$ modeling, unlike the rapidity distribution of the $W$ boson.
	We will focus on the predictions calculated at NLO unless specified.
	We find the PDF uncertainty of NNPDF3.1 tends to be similar in size comparing
	to the impact of varying $M_W$ by 5 MeV, and CT18 and MMHT14
	prefer a harder spectrum that is within the uncertainty of NNPDF3.1.
	That is in qualitative agreement with the CDF results concerning PDF
	uncertainty and dependence.
	The PDF variations of NNPDF4.0 are on the opposite side of CT18 and MMHT14,
	and are close to the boundaries of the uncertainty of NNPDF3.1.

	\begin{figure}
		\includegraphics[width=0.8\textwidth]{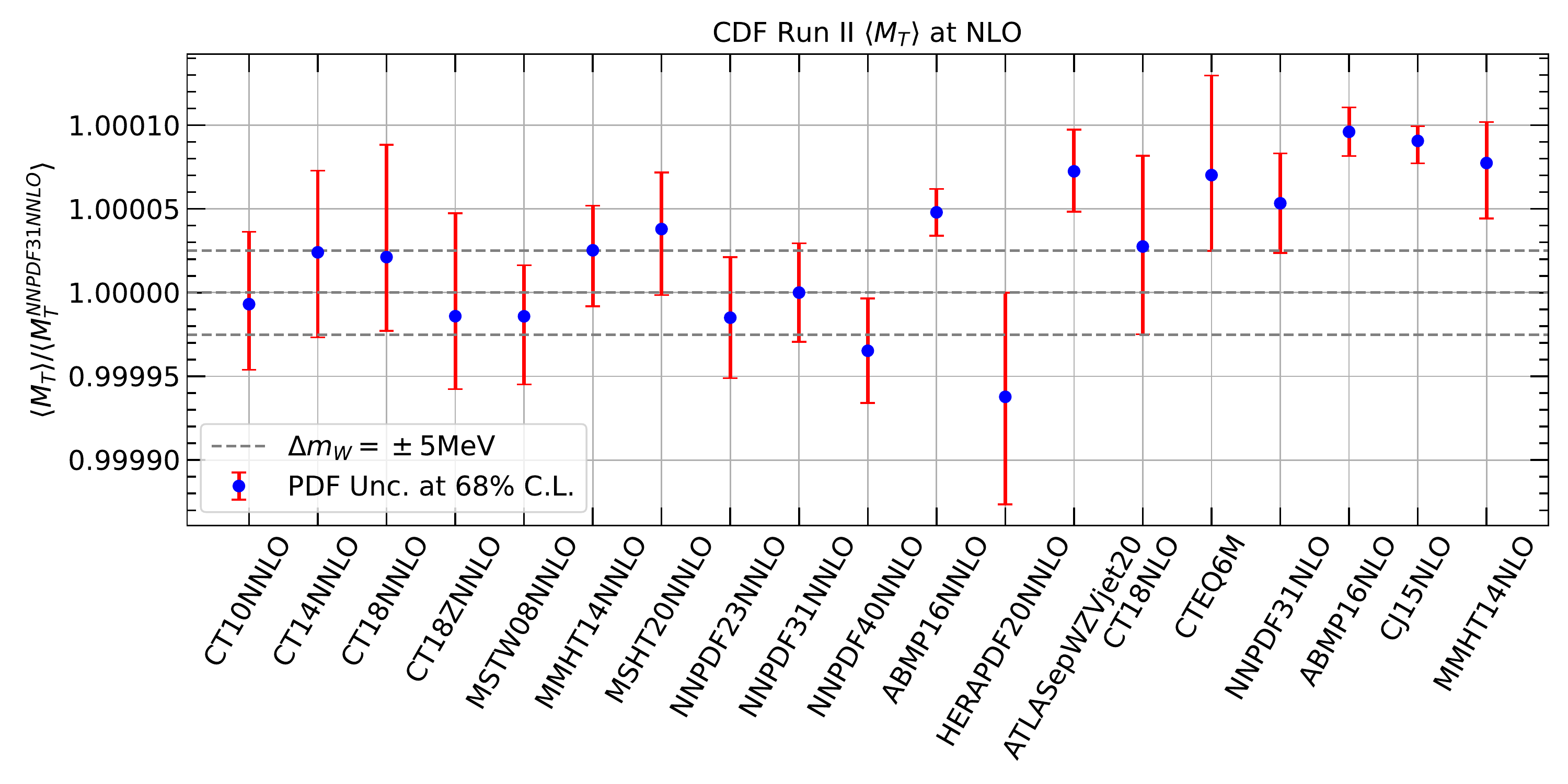}
		\includegraphics[width=0.8\textwidth]{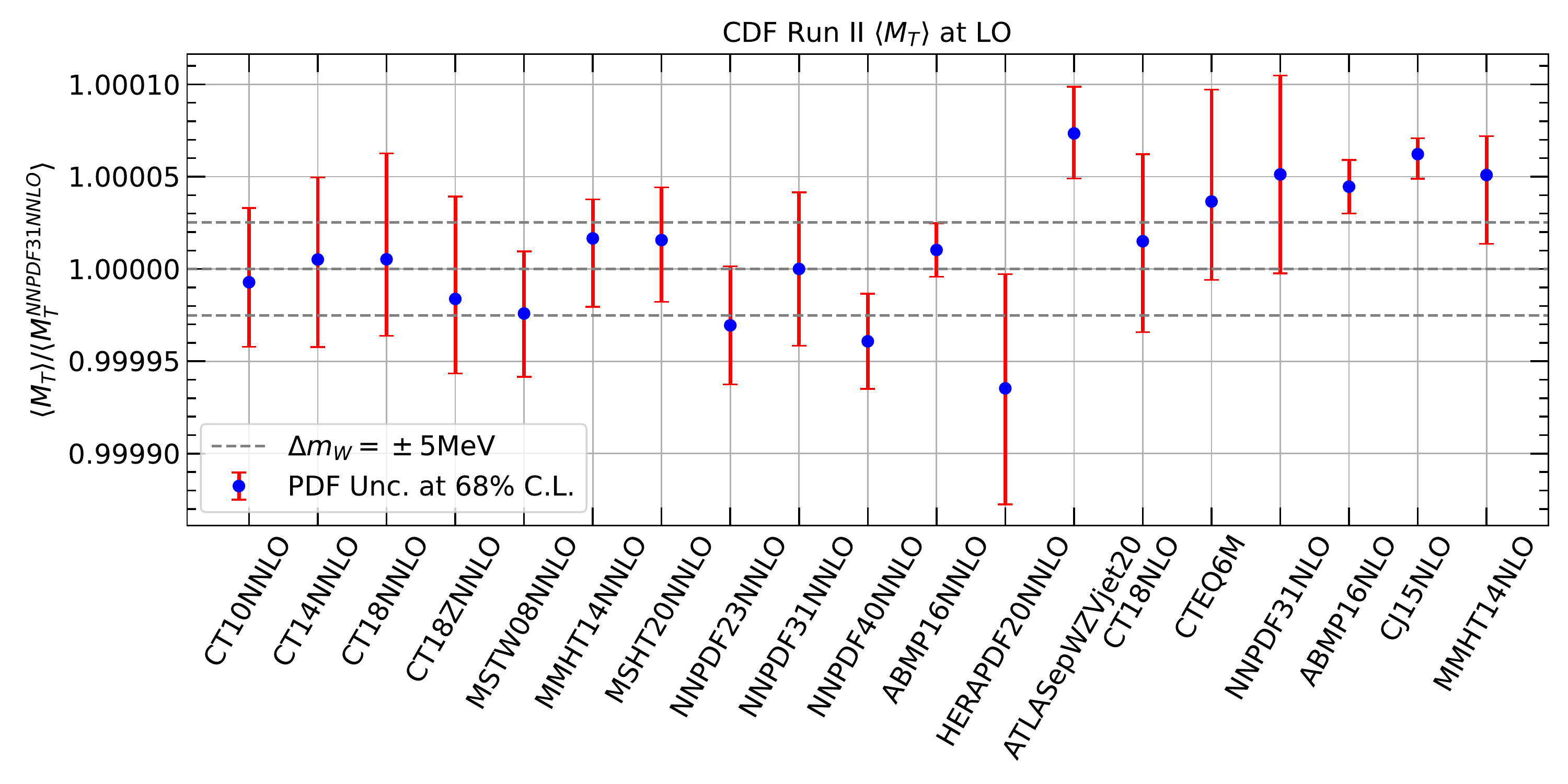}
		\caption{
			Mean transverse mass of the charged lepton and missing energies
			for the scenario of CDF measurement calculated at NLO or LO with various PDFs,
			normalized to the central prediction of NNPDF3.1 NNLO PDFs. 
			The error bars represent PDF uncertainties at 68\% C.L. and the horizontal
			lines indicate variations induced by a $W$ boson mass change of $\pm 5 $ MeV. 
		}
		\label{fig:cdfvar}
	\end{figure}

	We construct a principle variable to describe the impact of different PDFs
	on the shape of the $M_T$ distribution.
	That is the mean value of $M_T$ within a select window of [70, 90] GeV,
	denoted as $\langle M_T\rangle$.
	We choose this window in accordance with the CDF analysis, also because
	kinematic bins in this range show the largest significance when varying $M_W$.
	In Fig.~\ref{fig:cdfvar}, we plot various predictions on $\langle M_T\rangle$ at NLO and LO
	including PDF uncertainties at 68\% C.L. by normalizing to a common reference
	of the central prediction of NNPDF3.1.
	Here we further include the CT10~\cite{Gao:2013xoa}, CT14~\cite{Dulat:2015mca}, CT18Z~\cite{Hou:2019efy},
	MSTW2008~\cite{Martin:2009iq}, NNPDF2.3~\cite{Ball:2012cx}, MSHT20~\cite{Bailey:2020ooq}, ABMP16~\cite{Alekhin:2017kpj},
	HERAPDF2.0~\cite{H1:2015ubc}, ATLASepWZVjet20~\cite{ATLAS:2021qnl} and CJ15~\cite{Accardi:2016qay} PDFs
	for comparisons.
	We also show the range of $\langle M_T\rangle$ when varying $M_W$ by $\pm 5$ MeV.
	One can see predictions from all different NNLO PDFs agree within uncertainties
	in general.
	The spread of their central values at NLO can be as large as the shift of
	varying $M_W$ by 15 MeV if not taking into account those from HERA and ATLAS
	PDFs.
	The number is smaller and about 10 MeV for LO predictions.
	The size of PDF uncertainties is consistent among NNLO PDFs with 
	CT18 being the largest, about twice of that from NNPDF3.1.
	Besides, when comparing results using up-to-date PDFs to
	previous ones of the same PDF group, we find the PDF uncertainties
	can even increase in many cases.  
	The NLO PDFs predict larger $\langle M_T\rangle$ in general
	comparing to NNLO PDFs. 
	Fig.~\ref{fig:cdfvar} motivates a simplified prescription of using $\langle M_T\rangle$
	to quantify the shift of extracted $W$ boon mass when using different PDFs and the
	associated uncertainties.
	Ideally one can think of the shift of $M_W$ to compensate for the change of
	$\langle M_T\rangle$ induced by variation of PDFs.
	We summarize the estimated shift of extracted $M_W$ with respect to NNPDF3.1 and associated
	PDF uncertainties in Table.~\ref{tab:cdfmw}, and compare to those results in the CDF
	analysis when available.
	Using the $\langle M_T\rangle$ prescription NNPDF3.1 gives a PDF uncertainty of 5.7 MeV
	of the extracted $W$ boson mass at NLO, compared to a value of 3.9 MeV in the CDF
	analysis.
	The envelope of extracted $M_W$ values from central sets of NNPDF3.1, CT18, and MMHT14
	is about 5.0 MeV while the envelope is 4.2 MeV from the CDF analysis.
	Thus the simplified prescription is in a reasonable agreement with the dedicated
	simulation in the CDF analysis.
	However, using the $\langle M_T\rangle$ prescription we find a $\delta M_W$ value of
	-14 MeV comparing CTEQ6M with NNPDF3.1 while the value is -3.3 MeV from the CDF
	analysis. 
	The shift is reduced to -7.3 MeV when using LO calculations of $\langle M_T\rangle$.
	Another interesting observation is that for the most recent NNLO PDFs,
	namely CT18, MMHT20, and NNPDF4.0, the envelope of extracted $M_W$ from their central sets
	has been enlarged to 15(11) MeV at NLO(LO), larger than those of the previous generation.  
	That further motivates analyzing of the W boson mass data using up-to-date PDFs. 
	In the CDF analysis dependence of the extracted $M_W$ on PDFs are
	studied by fitting to pseudo-experiments.
	We perform similar studies assuming a pseudo-measurement of the $M_T$
	distribution equals our prediction of using NNPDF3.1 NNLO PDFs.
	We calculate the $\chi^2$ as a function of $M_W$ for using predictions from various
	PDFs, taking into account only statistical uncertainties since they are dominant
	over experimental systematic uncertainties.
	From there we can estimate the shift of extracted $M_W$ as well as the statistical and
	PDF uncertainties that are included in Table.~\ref{tab:cdfmw} for comparisons.
	We can see the $\chi^2$ fit indicates a statistical uncertainty on the extracted $M_W$
	of 8.0 MeV that is consistent with 9.2 MeV from the CDF measurement on the muon
	channel alone.
	The $\chi^2$ fit shows very good agreements on the projected shift
	of extracted $M_W$ for different PDFs with respect to previous simplified
	prescription of using $\langle M_T\rangle$.
	The estimated PDF uncertainties are slightly smaller than previous ones because the
	$\chi^2$ fit does account for higher moments of the kinematic distribution.

	\begin{table}[h!]
		\centering
		\begin{tabular}{c|cccccccc}
			\hline
			$\delta M_W$ in MeV & sta.& NNPDF3.1 &CT18& MMHT14& NNPDF4.0& MSHT20&CTEQ6M\\
			\hline
			$\langle M_T\rangle$(LO) &--&0$^{+ 8.3}_{-8.3}$ &$-$1.0$_{-11.4}^{+8.3}$&$-$3.3$_{-4.2}^{+7.4}$&$+$7.8$^{+5.1}_{-5.1}$&$-$3.1$_{-5.7}^{+6.7}$&$-$7.3$_{-12.0}^{+8.4}$ \\
			\hline
			$\chi^2$ fit (LO) &8.0& 0$_{-7.6}^{+7.6}$ &$-$1.0$_{-8.6}^{+5.4}$&$-$3.3$_{-3.0}^{+6.1}$&$+$8.0$_{-3.7}^{+3.7}$&$-$3.0$_{-4.0}^{+5.0}$&$-$7.3$_{-9.3}^{+5.6}$ \\         
			\hline
			$\langle M_T\rangle$(NLO) &--&0$^{+ 5.9}_{-5.9}$ &$-$4.2$_{-13.3}^{+8.8}$&$-$5.0$_{-5.3}^{+6.7}$&$+$6.9$^{+6.2}_{-6.2}$&$-$7.6$_{-6.7}^{+7.9}$&$-$14.0$_{-11.9}^{+9.0}$ \\
			\hline
			$\chi^2$ fit (NLO) &8.0& 0$_{-4.2}^{+4.2}$ &$-$4.3$_{-10.1}^{+5.4}$&$-$5.1$_{-3.4}^{+4.8}$&$+$7.1$_{-4.5}^{+4.5}$&$-$7.8$_{-4.5}^{+5.7}$&$-$14.6$_{-5.4}^{+5.8}$ \\
			\hline
			CDF &9.2& 0$^{+ 3.9}_{-3.9}$&--&--&--&--&$-$3.3 \\
			\hline
		\end{tabular}
		\caption{ %
			Estimated shifts and PDF uncertainties at 68\% C.L. on the extracted $W$ boson mass for the
			CDF scenario for various PDF sets with respect to a common reference of using NNPDF3.1 NNLO central PDF.
			We show results using the simplified prescription, compared to
			those from a $\chi^2$ fit as well as results reported in the CDF analysis. 
			In the case of the $\chi^2$ fit, we also show the expected experimental statistical error of the
			extracted $W$ boson mass compared to the actual one in the CDF analysis. 
		}
		\label{tab:cdfmw}
	\end{table}
	
	\begin{table}[]
		\centering
		\begin{tabular}{c|c|c|c}
			\hline
			Variation    & $\mu_{F,R}$ (7-point) & $\alpha_s=0.118\pm0.002$ & $\Gamma_W=2,085\pm42~\MeV$ \\
			\hline
			$\chi^2$ fit (NLO) & $0_{-3.0}^{+3.1}$ & $0_{-1.3}^{+1.2}$ & $0_{-6.8}^{+7.1}$  \\
			\hline
		\end{tabular}
		\caption{Dependence of $M_W$ extraction on the factorization and renormalization scales, the strong coupling and the $W$-boson decay width for the CDF scenario.}
		\label{tab:muasww}
	\end{table}

	Besides the PDF dependence, we have also explored other theoretical uncertainties,
	including the factorization and renormalization scales,
	the strong coupling constant and the $W$-boson decay width, 
	based on a $\chi^2$ fit of the transverse mass distribution at NLO with CT18 NNLO PDFs, 
	with results summarised in Tab.~\ref{tab:muasww}.  
	We discuss them in sequence as follows.\\
	$\bullet$
	The scale uncertainty is estimated with the envelope of 7-point variation,
	\begin{equation}
		(\mu_F,\mu_R)=\{(1/2,1/2),(1/2,1),(1,1/2),(1,1),(1,2),(2,1),(2,2)\}M_{\ell\nu}.
	\end{equation}
	It is found that maximal shifts on $M_W$ are -3.0 and +3.1 MeV, respectively.
	However, we expect the impact of scale variations to be largely reduced once higher order
	corrections are included.
	\\
	$\bullet$ 
	The strong coupling constant can impact $M_W$ extraction in two ways. 
	First, starting from NLO, high-order corrections to $W$ boson production directly involve
	in the QCD interaction. 
	Second, different choices of strong coupling in the QCD global analysis
	will lead to different PDFs, with impact propagating to the $M_W$ extraction. 
	We quantify the dependence by varying $\alpha_s$ between 0.116 and 0.120 together with consistent
	variations of PDFs.
	That changes the extracted $M_W$ by -1.3 and +1.2 MeV, respectively. \\
	$\bullet$ Currently, the most precise measurements of $W$-boson decay width 
	$\Gamma_W$ come from LEP~\cite{ALEPH:2013dgf} and Tevatron~\cite{TevatronElectroweakWorkingGroup:2010mao}, 
	which gave a combined result of $\Gamma_W=2,085\pm42~\MeV$~\cite{ParticleDataGroup:2020ssz}. 
	In experimental analysis, CDF collaboration adopted the electroweak
	global fitting value $2,089.5\pm0.6~\MeV$~\cite{ParticleDataGroup:2020ssz}, and found that
	uncertainty induced by the input width is negligible. 
	However, if allowing the $W$-boson width to vary by 42 MeV, the error from
	direct measurements, we found the normalized $M_T$ distribution can deviate
	significantly (up to 2\% around $M_T\sim90~\GeV$).
	As a result, the extracted $M_W$ value can shift by -6.8 (+7.1) MeV, comparable to the
	full uncertainty in the CDF measurement. 
	That suggests a simultaneous fit of the $W$-boson mass and width is possible and
	may result in comparable precision on the width measurement.

	\subsection{The ATLAS measurement}
\begin{figure}
		\includegraphics[width=0.49\textwidth]{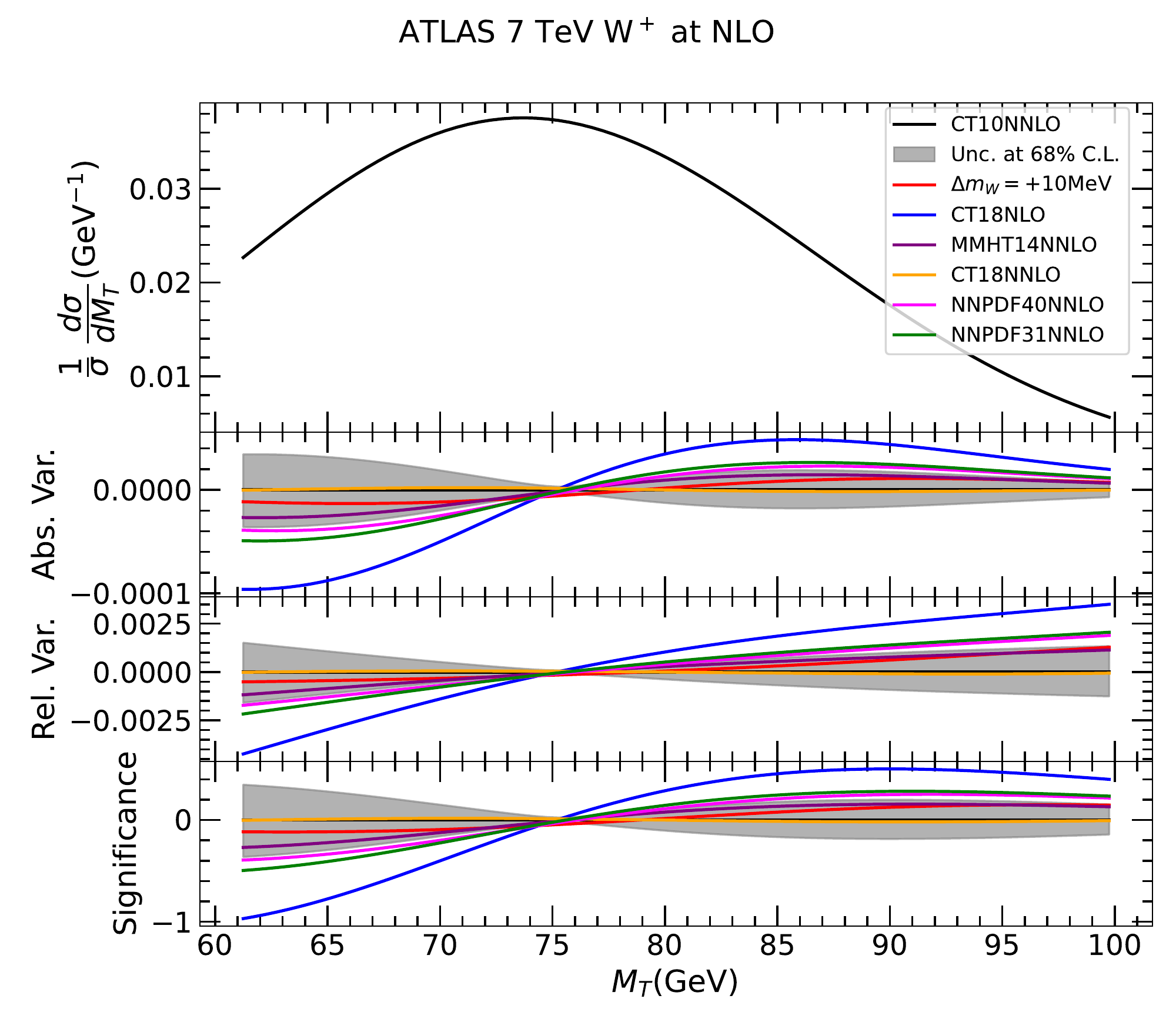}
		\includegraphics[width=0.49\textwidth]{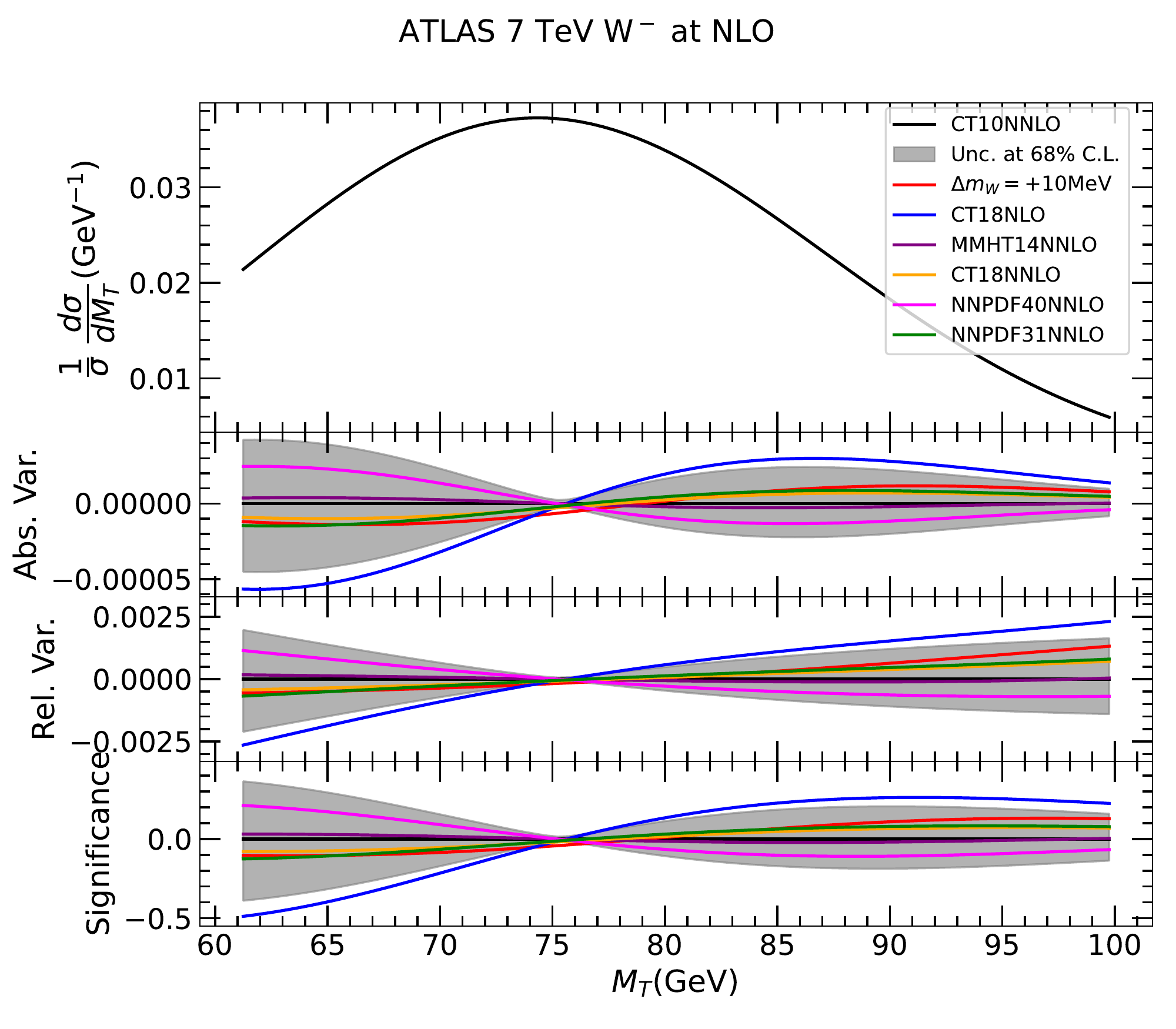}
		\caption{
			Transverse mass distribution of the charged lepton and missing energies
			for the scenario of the ATLAS measurement for $W^+$ and $W^-$ with various PDFs
			and different choices of the $W$ boson mass (increased by 10 MeV) calculated at NLO. 
			From top to bottom 
			are the normalized distribution, absolute and relative changes with respect
			to a common reference of the prediction
			obtained with CT10 NNLO PDFs and nominal $W$ boson mass.
			The lowest panel shows the changes normalized to the experimental
			statistical uncertainties.
		}
		\label{fig:atldis}
	\end{figure}

	We repeat a similar exercise for the ATLAS 7 TeV measurement.
	The event selection criterion follows~\cite{ATLAS:2017rzl}
	\begin{equation}
		p_T^{\ell,\nu}>30~\GeV,~u_T<30~\GeV,~ M_T>60~\GeV.
	\end{equation}
	The pseudo-rapidity of the charged lepton is required to satisfy
	\begin{equation}
		|\eta|<2.4.
	\end{equation}
	We use the same theoretical setups as in the calculations for the CDF scenario
	except that we need to calculate separately for the $W^+$ and $W^-$ production.
	We assume a detector resolution of 10\% on $M_T$ in order to reproduce well the shape of the measured distribution.
	In Fig.~\ref{fig:atldis}, we show the predictions on the normalized $M_T$ distribution
	at NLO for several choices of the PDFs and with different choices of the $W$ boson mass,
	for both the $W^+$ and $W^-$ production.
	We find the PDF variations are about twice of those shown in Fig.~\ref{fig:cdfdis}
	for the CDF scenario while the dependence on the $W$ boson mass is similar in size.
	The PDFs clearly alter the $W^+$ and $W^-$ production in different ways as evident
	that the NNPDF4.0 central predictions lie on opposite sides of CT10 for $W^+$ and $W^-$.
	The PDF uncertainties of CT10 are larger for $W^-$ production than $W^+$ production
	possibly due to the relatively larger contributions from the strange quark in the former case.
	The significance in the lowest panel is calculated assuming a total number of events
	equals to that from the ATLAS measurement of the muon channel and with a bin width of
	0.5 GeV.

\begin{figure}
		\includegraphics[width=0.8\textwidth]{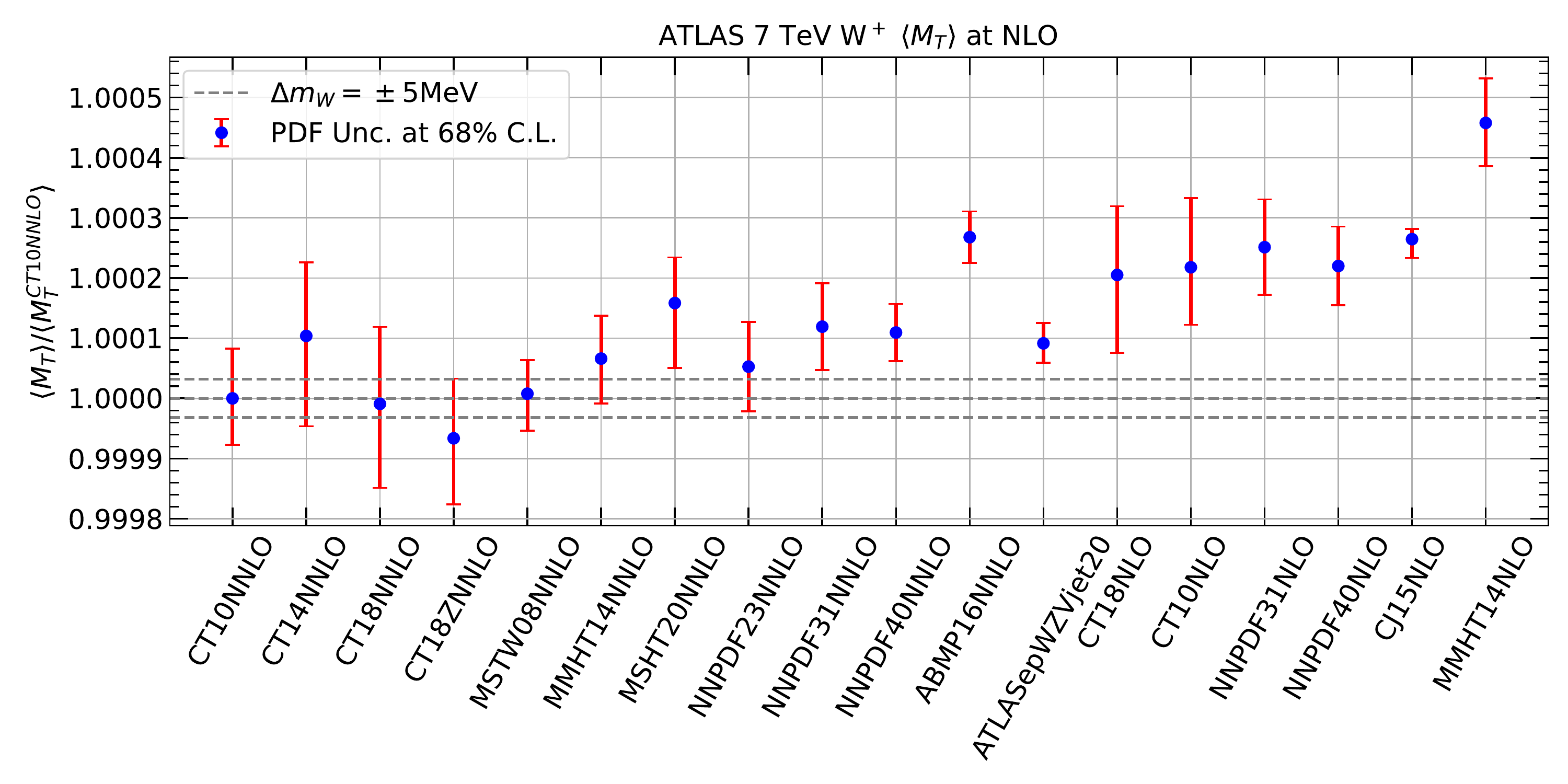}
		\includegraphics[width=0.8\textwidth]{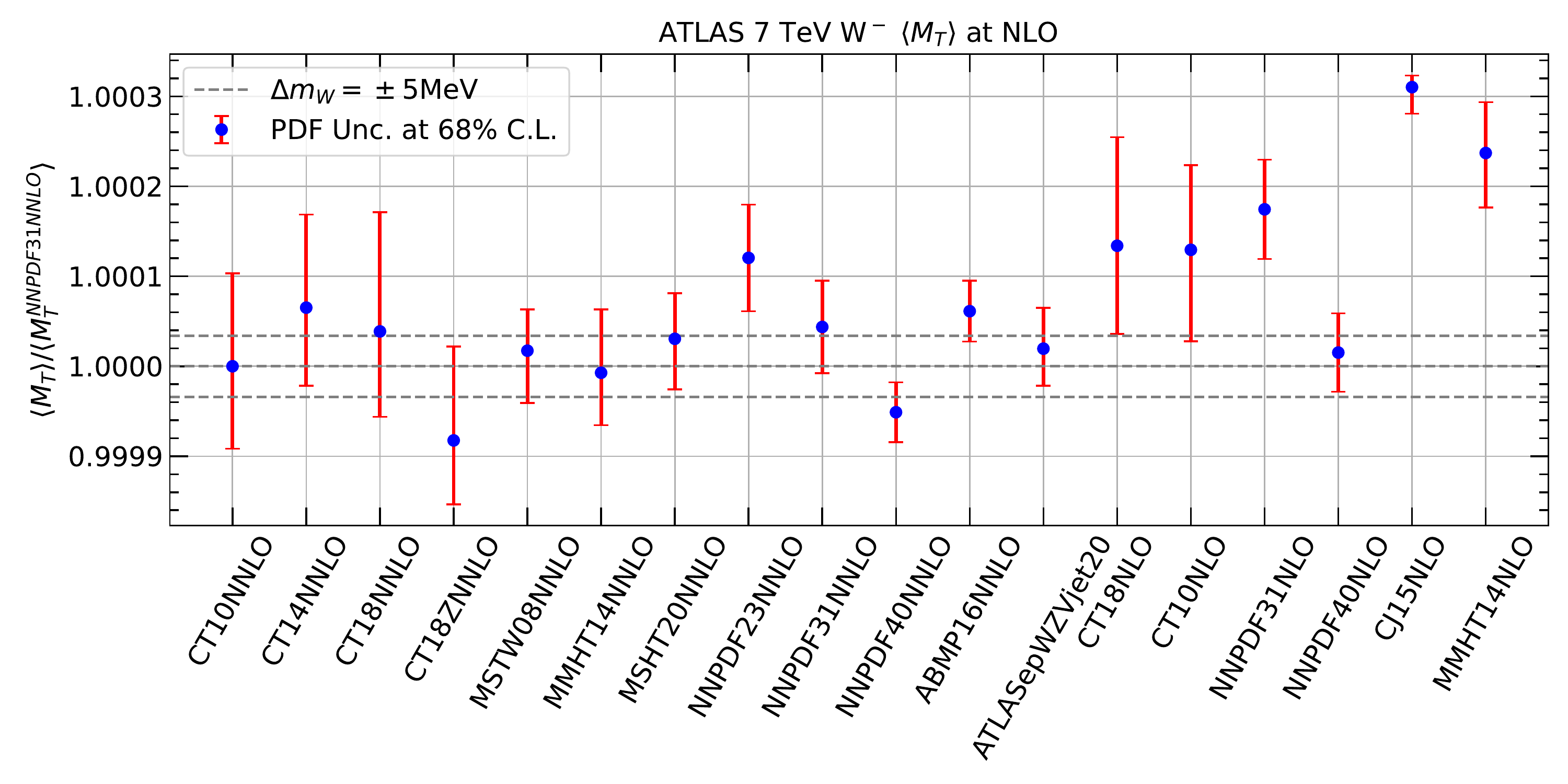}
		\caption{
			Mean transverse mass of the charged lepton and missing energies
			for the scenario of ATLAS measurement calculated at NLO with various PDFs,
			normalized to the central prediction of CT10 NNLO PDFs, for $W^+$
			and $W^-$ production respectively. 
			The error bars represent PDF uncertainties at 68\% C.L. and the horizontal
			lines indicate variations induced by a $W$ boson mass change of $\pm 5 $ MeV. 
		}
		\label{fig:atlvar}
	\end{figure}

	Similar to the CDF case, we now plot in Fig.~\ref{fig:atlvar} the mean transverse mass
	within a window of [65, 100] GeV for the $W^+$ and $W^-$ production respectively, for
	various choices of PDFs.
	All predictions are normalized to the central prediction from CT10 NNLO PDFs including those
	with a $W$ boson mass change of $\pm 5$ MeV.
	We find a wider spread of predictions from different PDFs and larger PDF uncertainties
	comparing to Fig.~\ref{fig:cdfvar}, consistent with observations in the normalized
	distributions.
	We can again use the simplified prescription to estimate the expected shift on
	the extracted $W$ boson mass and the associated PDF uncertainties, which are summarized
	in Table.~\ref{tab:atlmw}.
	For example, we estimate a PDF uncertainty of about 13 and 14 MeV with CT10 NNLO PDFs
	for $W^+$ and $W^-$ respectively, while the ATLAS report 15 and 14 MeV. 
	We also note that both the central sets of NNPDF4.0 and MSHT20 prefer a downward shift
	of $M_W$ of almost 20 MeV for the $W^+$ production compared to CT10.
	The shifts are generally smaller for the $W^-$ production.
	We take an unweighted average of the $W^+$ and $W^-$ results on $\langle M_T\rangle$
	and show the shift of the extracted $W$ boson mass in Table.~\ref{tab:atlmw} denoted as $W^{\pm}$. 
	The combinations show less spread on the expected shift of $M_W$, and the PDF uncertainties
	are in general reduced. 
	The PDF uncertainty on the $W$ boson mass for $W^{\pm}$ goes down slightly to 11 MeV for
	CT10 NNLO PDFs.
	We notice in the ATLAS analysis the combined $W^{\pm}$ results show a PDF uncertainty
	of about 7 MeV, largely reduced compared to those of $W^+$ and $W^-$ alone as
	been explained due to the anti-correlations of PDF dependence in the two.
	We do observe a similar pattern for CT10 if using LO calculations instead of NLO,
	namely excluding PDF uncertainties due to the modeling of the $W$ boson $p_T$. 
	The corresponding results are also summarized in Table.~\ref{tab:atlmw}.

	\begin{table}[h!]
		\centering
		\begin{tabular}{c|cccccc}
			\hline
			$\delta M_W$ in MeV & CT10 &CT18& MMHT14& NNPDF4.0& CT14 &MSHT20\\
			\hline
			$W^+$ $\langle M_T\rangle$ (NLO) &0$^{+ 12.1}_{-12.9}$ &$+$1.4$_{-20.0}^{+21.8}$&$-$10.3$_{-11.1}^{+11.6}$&$-$17.1$^{+7.4}_{-7.4}$&$-$16.2$_{-19.1}^{+23.5}$&$-$24.8$_{-11.9}^{+16.8}$ \\
			\hline
			$W^-$ $\langle M_T\rangle$ (NLO) &0$^{+ 13.5}_{-15.2}$ &$-$5.7$_{-19.5}^{+14.0}$&$+$1.1$_{-10.3}^{+8.6}$&$+$7.5$^{+4.9}_{-4.9}$&$-$9.6$_{-15.3}^{+12.8}$&$-$4.5$_{-7.5}^{+8.3}$ \\
			\hline
			$W^{\pm}$ $\langle M_T\rangle$ (NLO) &0$^{+ 9.8}_{-11.4}$ &$-$2.3$_{-16.8}^{+14.4}$&$-$4.5$_{-8.5}^{+8.2}$&$-$4.4$^{+4.6}_{-4.6}$&$-$12.8$_{-15.1}^{+16.6}$&$-$14.3$_{-8.0}^{+10.9}$ \\
			\hline
			$W^+$ $\langle M_T\rangle$ (LO) &0$^{+ 10.8}_{-11.4}$ &$-$6.5$_{-10.0}^{+14.1}$&$-$5.7$_{-7.1}^{+8.1}$&$-$14.1$^{+5.8}_{-5.8}$&$-$4.1$_{-12.9}^{+15.0}$&$-$14.4$_{-7.3}^{+10.2}$ \\
			\hline
			$W^-$ $\langle M_T\rangle$ (LO) &0$^{+ 8.9}_{-11.4}$ &$-$7.2$_{-12.5}^{+10.1}$&$+$3.1$_{-9.9}^{+8.3}$&$+$3.5$^{+4.5}_{-4.5}$&$-$7.0$_{-8.9}^{+6.2}$&$+$2.1$_{-4.9}^{+6.3}$ \\
			\hline
			$W^{\pm}$ $\langle M_T\rangle$ (LO) &0$^{+ 5.2}_{-7.0}$ &$-$0.6$_{-7.4}^{+7.6}$&$-$1.2$_{-5.9}^{+5.3}$&$-$5.0$^{+3.0}_{-3.0}$&$-$5.6$_{-8.4}^{+8.0}$&$-$5.9$_{-4.2}^{+5.9}$ \\  
			\hline
			$W^+$ ATLAS & 0$^{+ 14.9}_{-14.9}$&--&--&--&--&-- \\
			\hline
			$W^-$ ATLAS & 0$^{+ 14.2}_{-14.2}$&--&--&--&--&-- \\
			\hline
			$W^{\pm}$ ATLAS & 0$^{+ 7.4}_{-7.4}$&--&--&--&--&-- \\
			\hline
		\end{tabular}
		\caption{
			Estimated shifts and PDF uncertainties at 68\% C.L. of the extracted $W$ boson mass for the
			ATLAS scenario for various PDF sets with respect to a common reference of using CT10 NNLO central PDF.
			We show results using the simplified prescription and calculations at NLO and LO, and
			compare to the numbers in the ATLAS analysis. 
		}
		\label{tab:atlmw}
	\end{table}

	\section{Lagrange multiplier scan}~\label{sec:B}

	In this section, we are dedicated to studying PDF uncertainties of the $W$ boson mass measurements
	in the context of the CT18 global analysis.
	We focus on the observable of the mean transverse mass that is strongly anti-correlated with the
	extracted $W$ boson mass as been shown.
	We first show correlations between $\langle M_T\rangle$ and PDFs of different flavors
	and at different $x$ values, and of $\langle M_T\rangle$ in different measurements.
	That is followed by a series of Lagrange multiplier scans on understanding constraints
	as imposed by individual data sets in the CT18 global analysis.

	\subsection{PDF induced correlations}

\begin{figure}
		\includegraphics[width=0.49\textwidth]{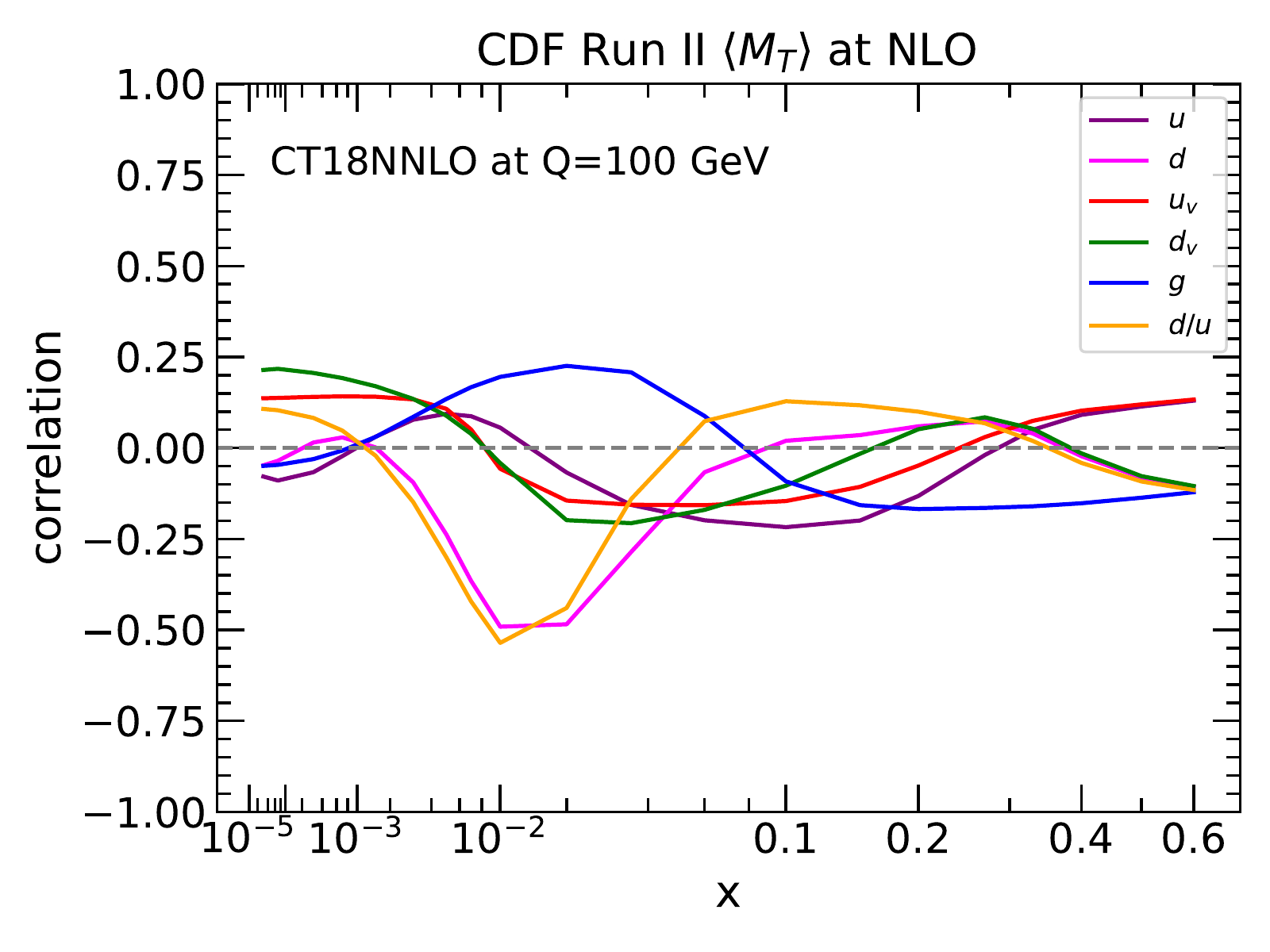}
		\includegraphics[width=0.49\textwidth]{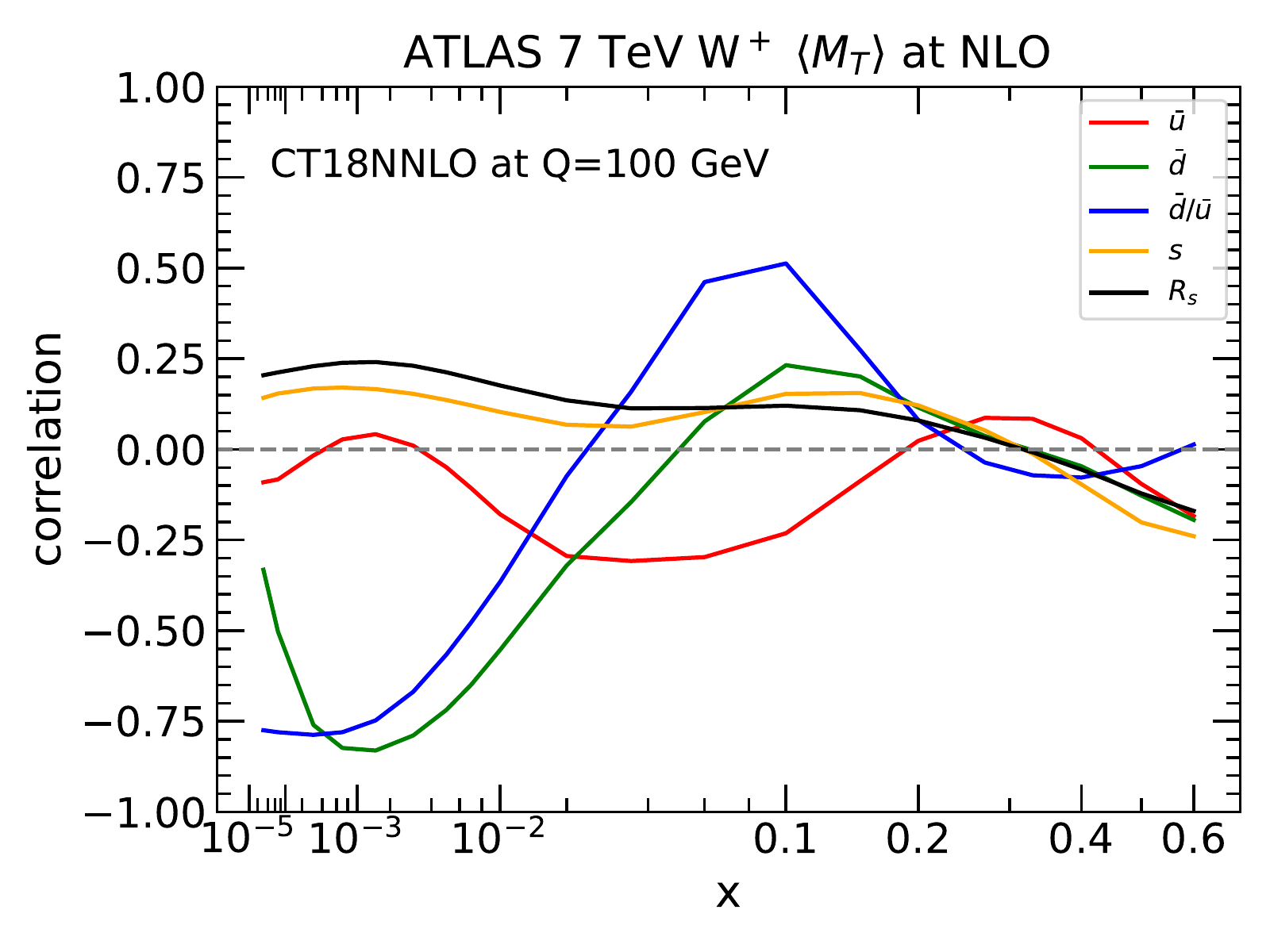}
		\includegraphics[width=0.49\textwidth]{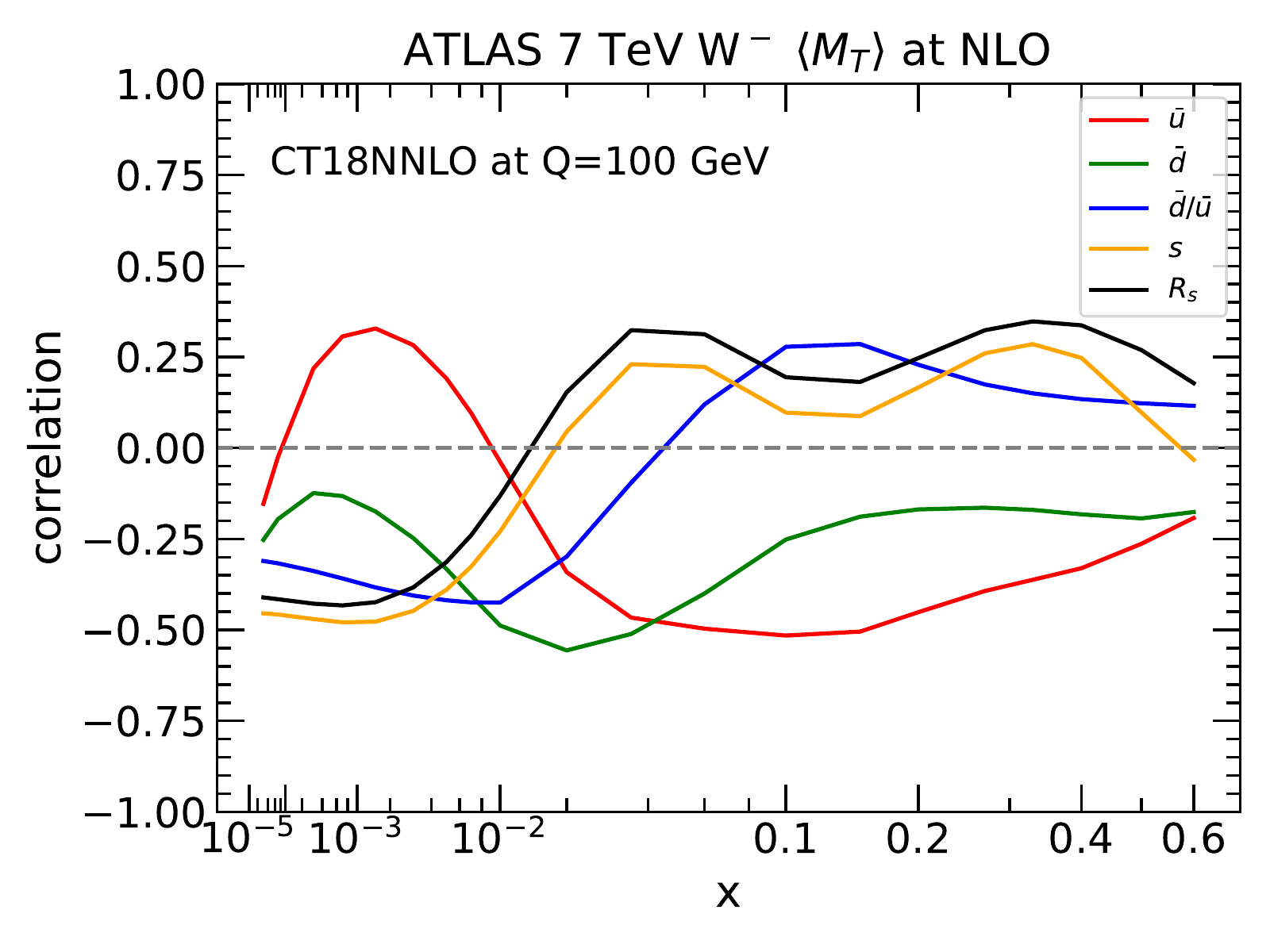}
		\includegraphics[width=0.49\textwidth]{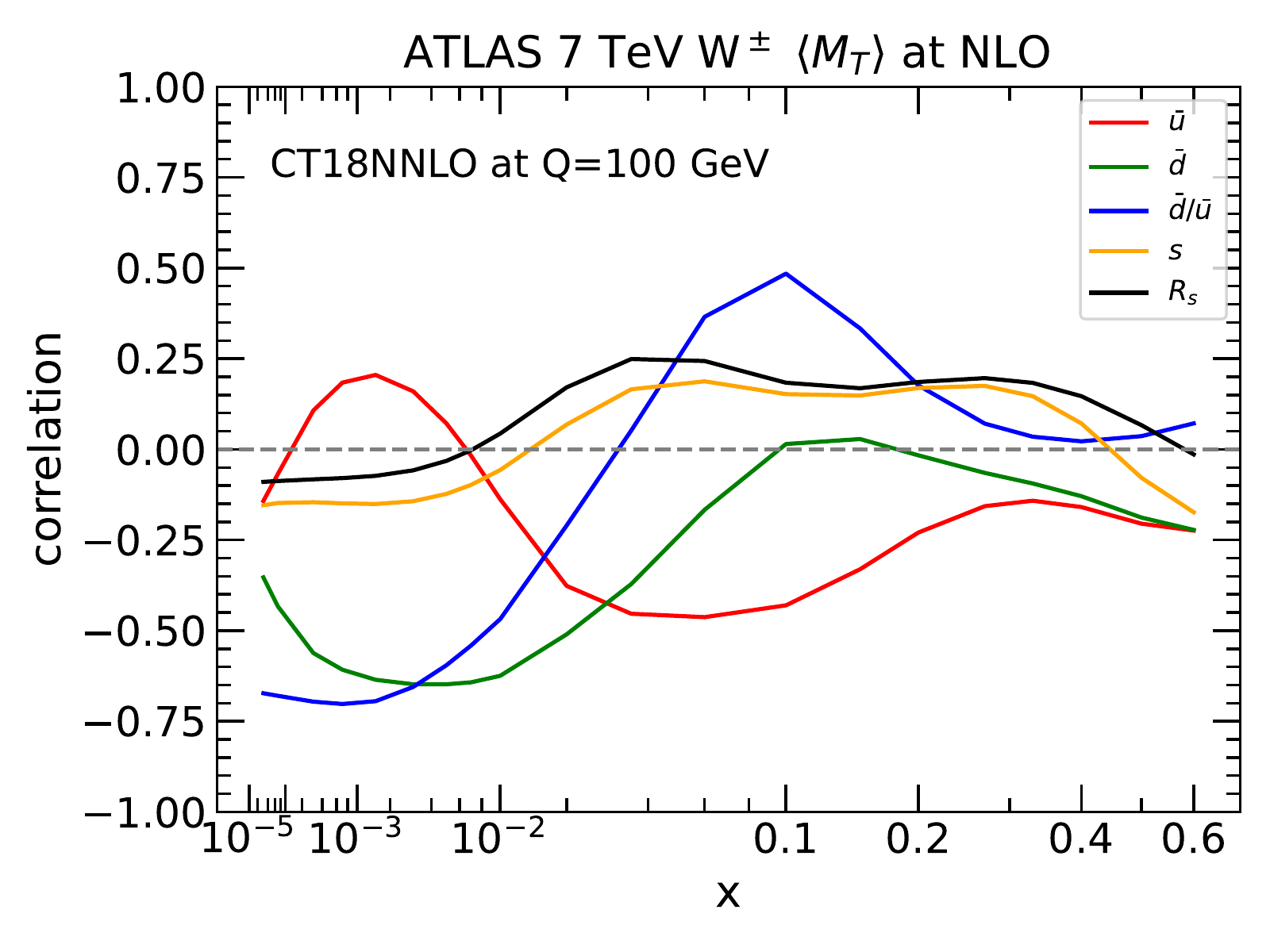}
		\caption{
			Correlations between $\langle M_T\rangle$ calculated at NLO and
			PDFs of different flavors as a function of $x$ for the scenario of
			the CDF measurement and the ATLAS measurements 
			of $W^+$, $W^-$ and combined, using CT18 NNLO PDFs. 
		}
		\label{fig:cor1}
	\end{figure}

	We study the PDF induced correlations of the observables
	proposed in the last section, namely the mean transverse mass of the charged
	lepton and missing energies in the CDF and the ATLAS measurements.
	The correlations are calculated using CT18 NNLO PDFs and with the transverse mass
	distributions at NLO by default.
	In Fig.~\ref{fig:cor1} we plot the correlations between $\langle M_T\rangle$
	and the PDF of various flavors at different $x$ values and with $Q=$ 100 GeV.
	For the scenario of the CDF measurement, $\langle M_T\rangle$ is anti-correlated
	with $d$-quark at $x\sim 0.01$ since that corresponds to a $W$ boson produced
	in large rapidity regions where the decayed lepton has a smaller average $p_T$,
	as will be explained later in Sec.~\ref{sec:dis}.
	For the same reason $\langle M_T\rangle$ in the case of the $W^+$ production at
	ATLAS 7 TeV is anti-correlated with the $\bar d$ quark but now at $x\sim 0.002$.
	In the $W^-$ production at ATLAS 7 TeV, various sea quarks show moderate
	anti-correlations including the strange quark.
	For the average $\langle M_T\rangle$ of $W^+$ and $W^-$ production at ATLAS,
	the correlations show an average pattern between the two.

\begin{figure}
		\includegraphics[width=0.45\textwidth]{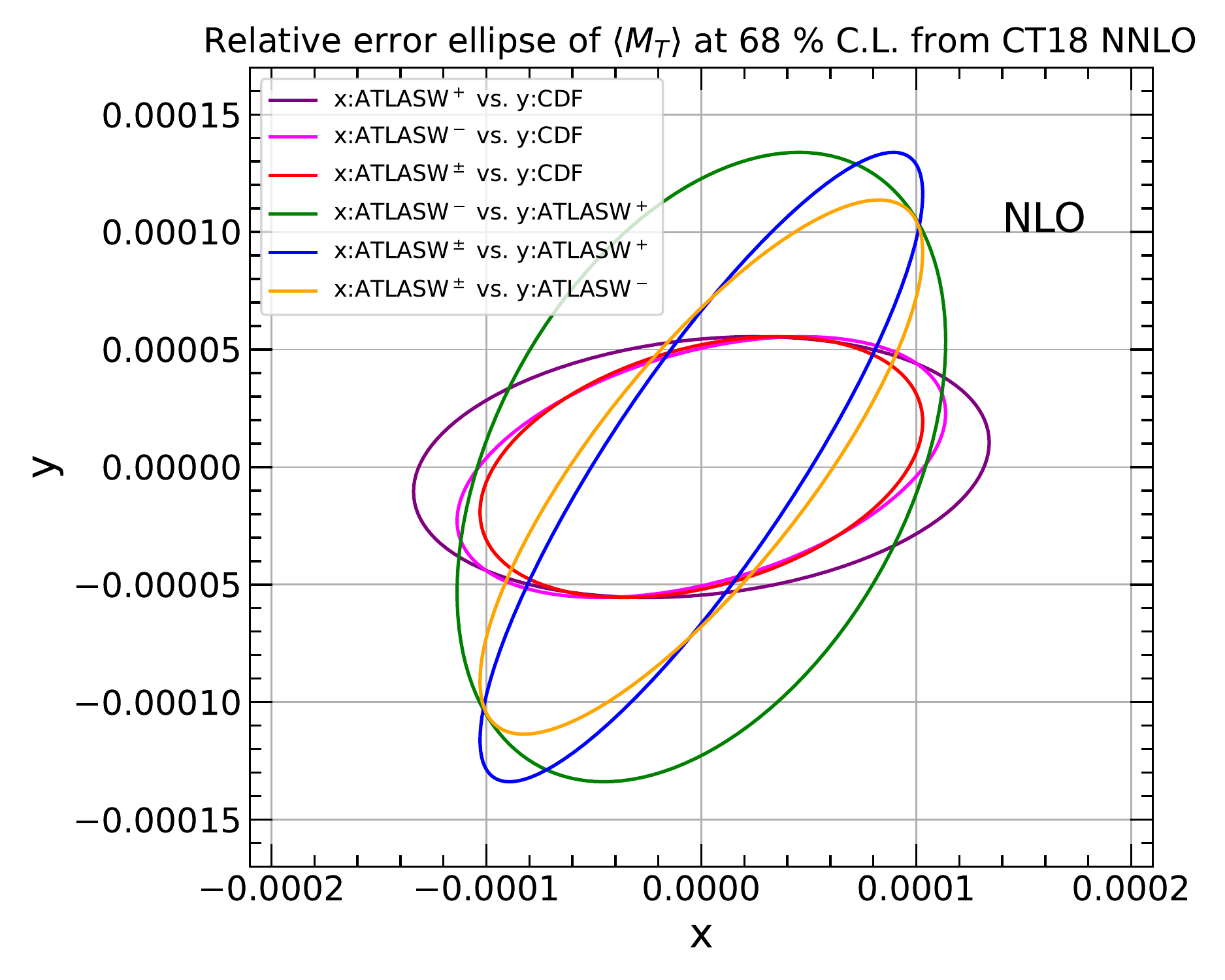}
		\includegraphics[width=0.45\textwidth]{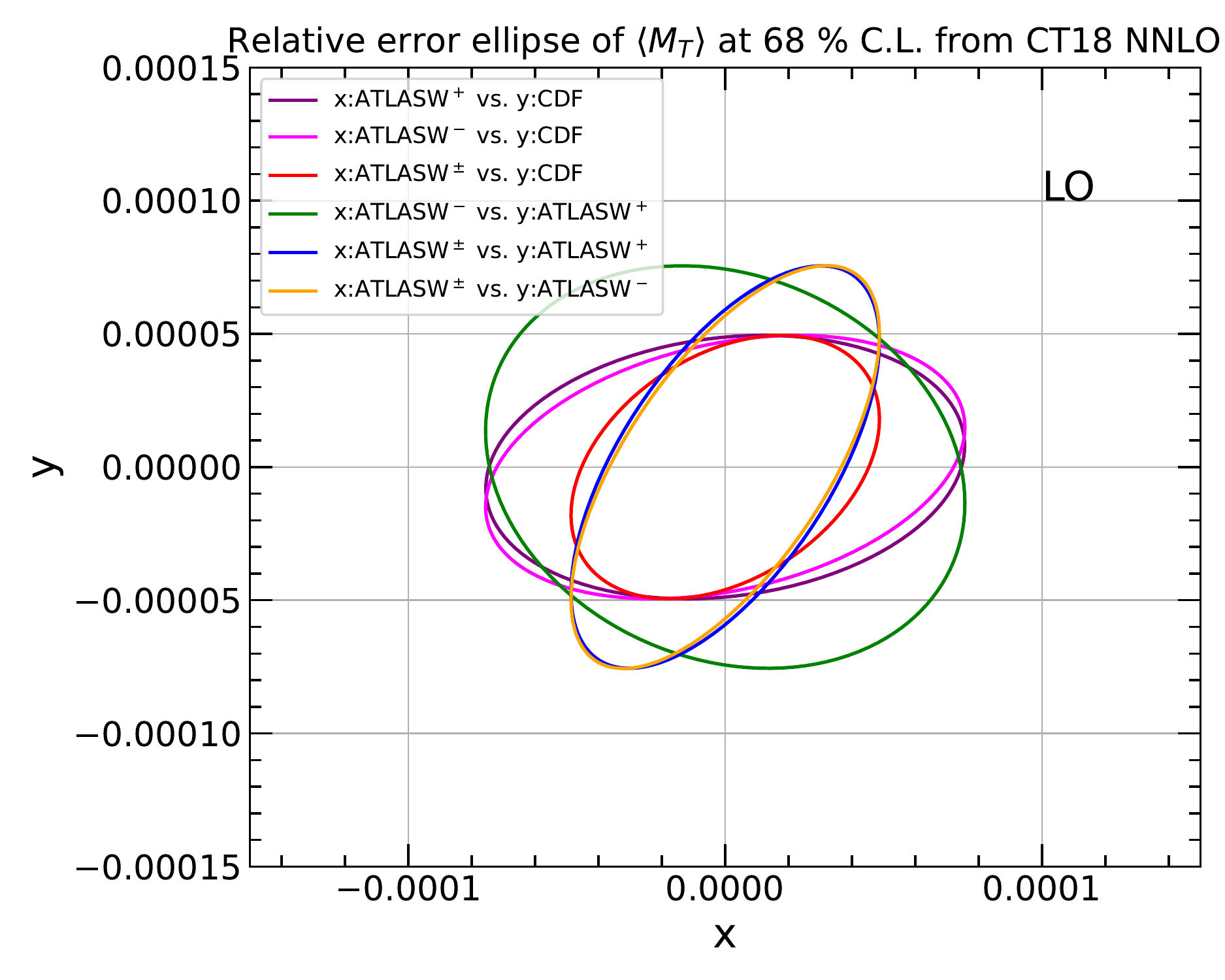}
		\caption{
			Relative error ellipse at 68\% C.L. for each pair of $\langle M_T\rangle$ for the scenario of
			the CDF measurement and the ATLAS measurements of $W^+$, $W^-$ and combined,
			calculated at NLO and LO using CT18 NNLO PDFs. 
		}
		\label{fig:cor2}
	\end{figure}

	We plot the 68\% C.L. relative error ellipse for each pair of the observables in
	Fig.~\ref{fig:cor2} at both NLO and LO.
	One can see the PDF uncertainty of $\langle M_T\rangle$ at ATLAS is about
	twice of that at CDF, which is consistent with results shown in an earlier section,
	and also the two are largely uncorrelated.
	That means the PDF uncertainty on the extracted $M_W$ can be further
	reduced when combining the CDF and the ATLAS measurements.
	On another hand, the PDF uncertainties of $\langle M_T\rangle$ for $W^+$
	and $W^-$ production at the ATLAS are only partially correlated at the NLO
	or anti-correlated at the LO.
	Combining $\langle M_T\rangle$ of $W^+$ and $W^-$ will reduce
	the PDF uncertainty especially at the LO as it is evident from
	Table.~\ref{tab:atlmw}.

	\subsection{Constraints in CT18}
	LM scan is a robust method to estimate PDF uncertainties, which was originally developed in Refs.~\cite{Pumplin:2000vx,Stump:2001gu}.
	In this method, PDF uncertainties of an observable can be determined from the profiled $\chi^2$ as a function of the observable, without relying on any assumptions about the specific behavior of the $\chi^2$ around
	the global minimum.
	However, the LM method requires a detailed scan of the PDF parameter space for every observable studied, which is usually time-consuming.
	To overcome this drawback, we take advantage of Neural Networks (NNs) and machine learning techniques
	to model profile of the $\chi^2$ and $\langle M_{T} \rangle$ for multi-dimensional parameter space,
	which works beyond the quadratic approximations and meanwhile ensures efficient scans of the full parameter space.
	The setup of the NNs and further details can be found in Ref.~\cite{Liu:2022plj}.

\begin{figure}
		\includegraphics[width=0.49\textwidth]{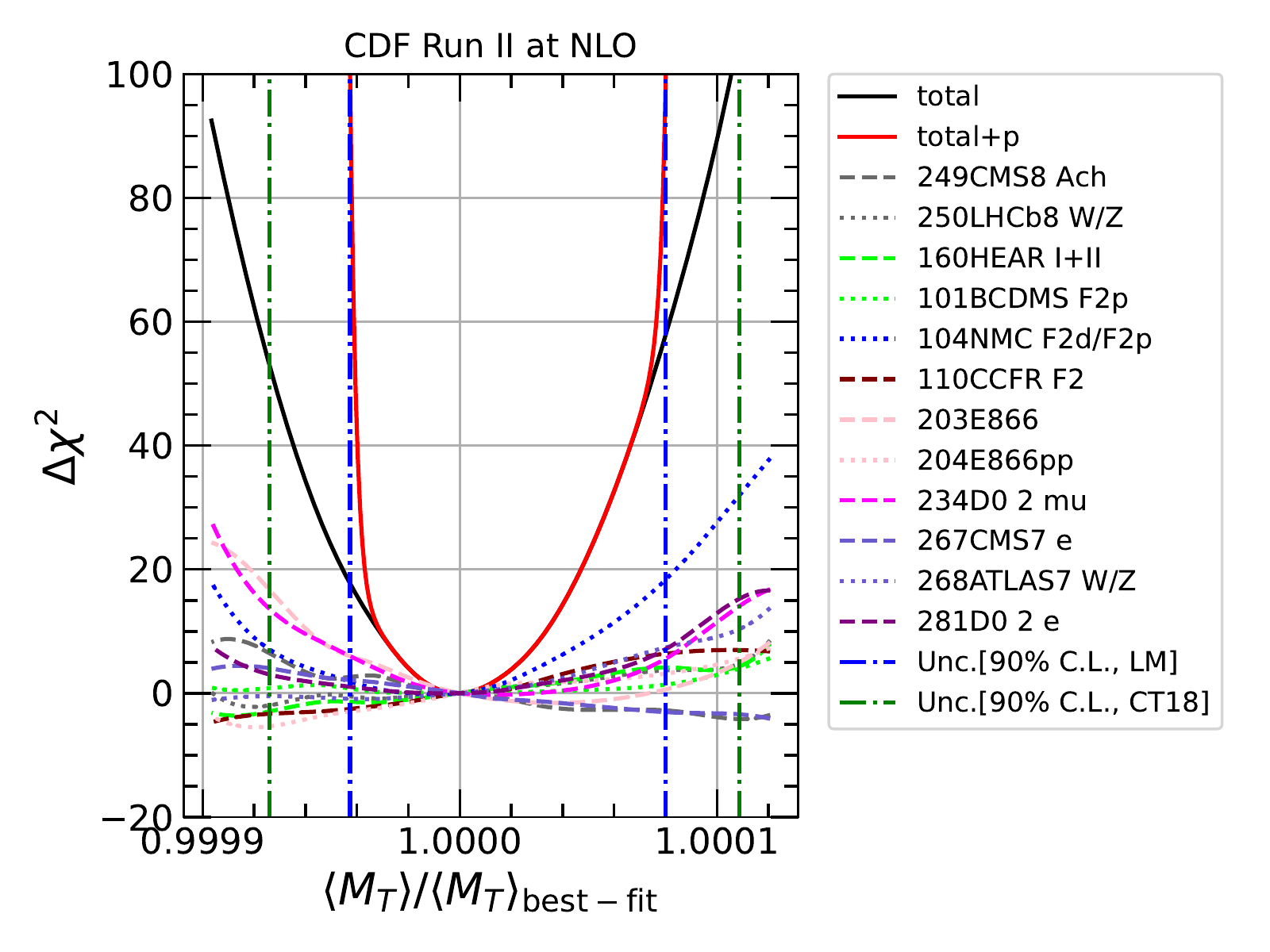}
		\includegraphics[width=0.49\textwidth]{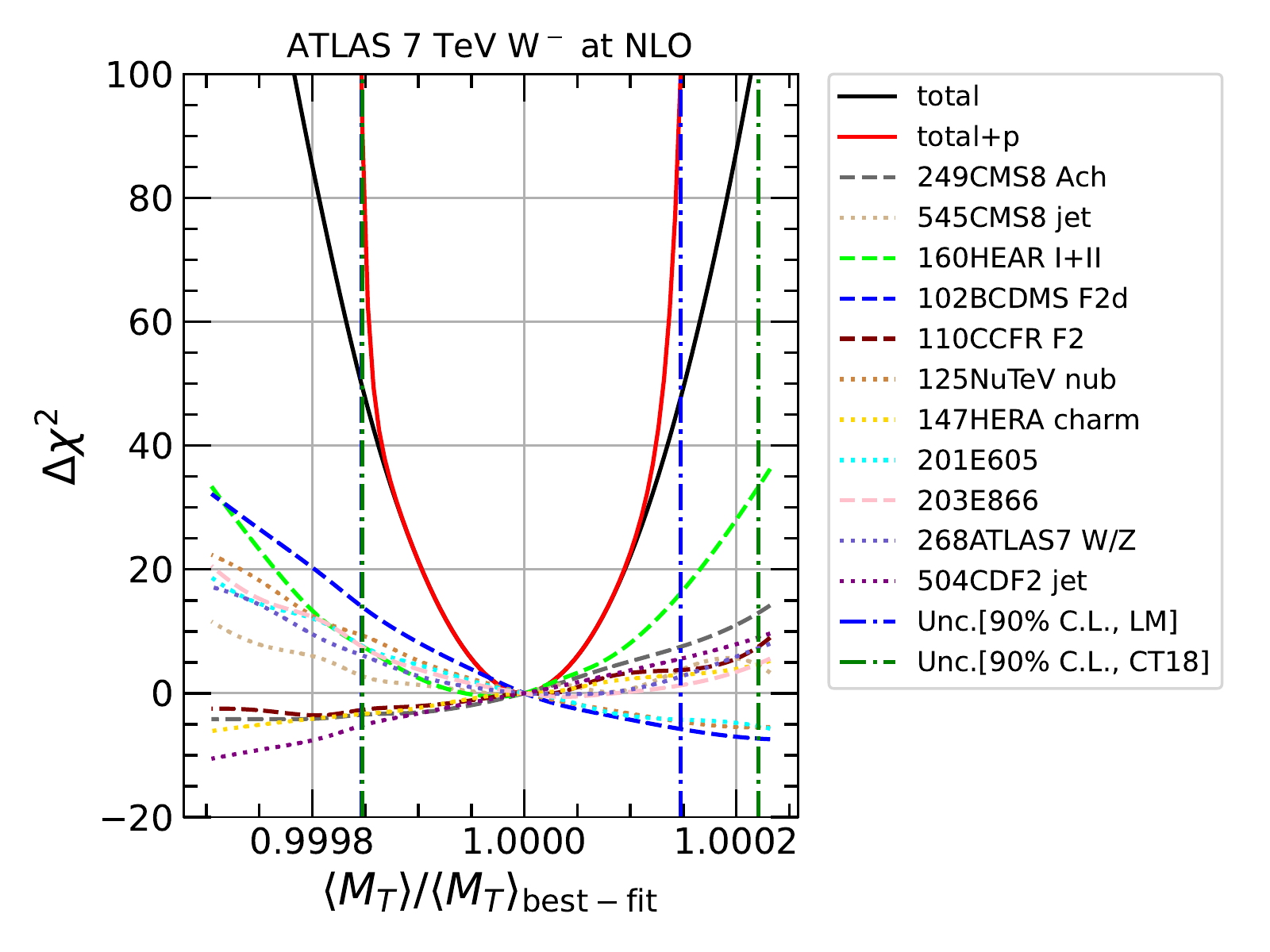}
		\caption{LM scans on $\langle M_T\rangle$.
			The black and the red solid line represent $\Delta\chi^2$ and $\Delta\chi^2 + P$ respectively.
			The dot and the dash lines indicate the contributions to $\Delta\chi^2$ from individual data sets.
			The blue and the green vertical dot-dash lines represent the uncertainties at 90\% C.L. determined with the LM method and with the Hessian method from the published CT18 NNLO PDFs, respectively.}
		\label{fig:LM_mt}
	\end{figure}

	In Fig.~\ref{fig:LM_mt} we show the results of LM scans on $\langle M_{T}\rangle$ based on the aforementioned NNs. 
	The black and the red solid line indicate $\Delta\chi^2$ and $\Delta\chi^2 + P$ respectively, where $P$, called Tier-2 penalty~\cite{Gao:2017yyd,Dulat:2015mca}, is introduced to ensure the tolerance will be reached as soon as any data set shows disagreement at 90\% C.L.
	The PDF uncertainty of $\langle M_{T}\rangle$ at 90\% C.L. can be determined by requiring $\Delta\chi^2 + P = 100$
	following the CT18 analysis.
	The dot and the dash lines represent the contributions to $\Delta\chi^2$ from individual data sets.
	The blue and the green vertical dot-dash lines indicate the uncertainties at 90\% C.L. determined
	with the LM method and with the Hessian method from the published CT18 NNLO PDFs, respectively.
	It can be seen that the profile of the total $\Delta\chi^2$ and individual $\Delta\chi^2$ show almost a quadratic dependence on the variable at the neighborhood of the global minimum.
	For the case of the CDF measurement, in the left panel, the NMC deuteron to proton ratio data together with the D0 Run II charge asymmetry data and the E866 Drell-Yan deuteron to proton ratio data give the dominant constraints. 
	The penalty term contributes largely to the total constraints and cuts off the
	uncertainty range well before the global $\Delta \chi^2$ reaches the tolerance.
	The LM method gives a smaller PDF uncertainty than the estimation based on the Hessian method.
	For the case of the $W^{-}$ production at ATLAS 7 TeV, in the right panel, we find that the BCDMS deuteron data together with the HERA inclusive DIS data and the CMS 8 TeV charge asymmetry data give the dominant constraints.
	In addition, the BCDMS deuteron data prefer a larger $\langle M_{T}\rangle$, which results in a large penalty term.
	The LM method again predicts a slightly smaller uncertainty than the Hessian method.

\begin{figure}
		\includegraphics[width=0.7\textwidth]{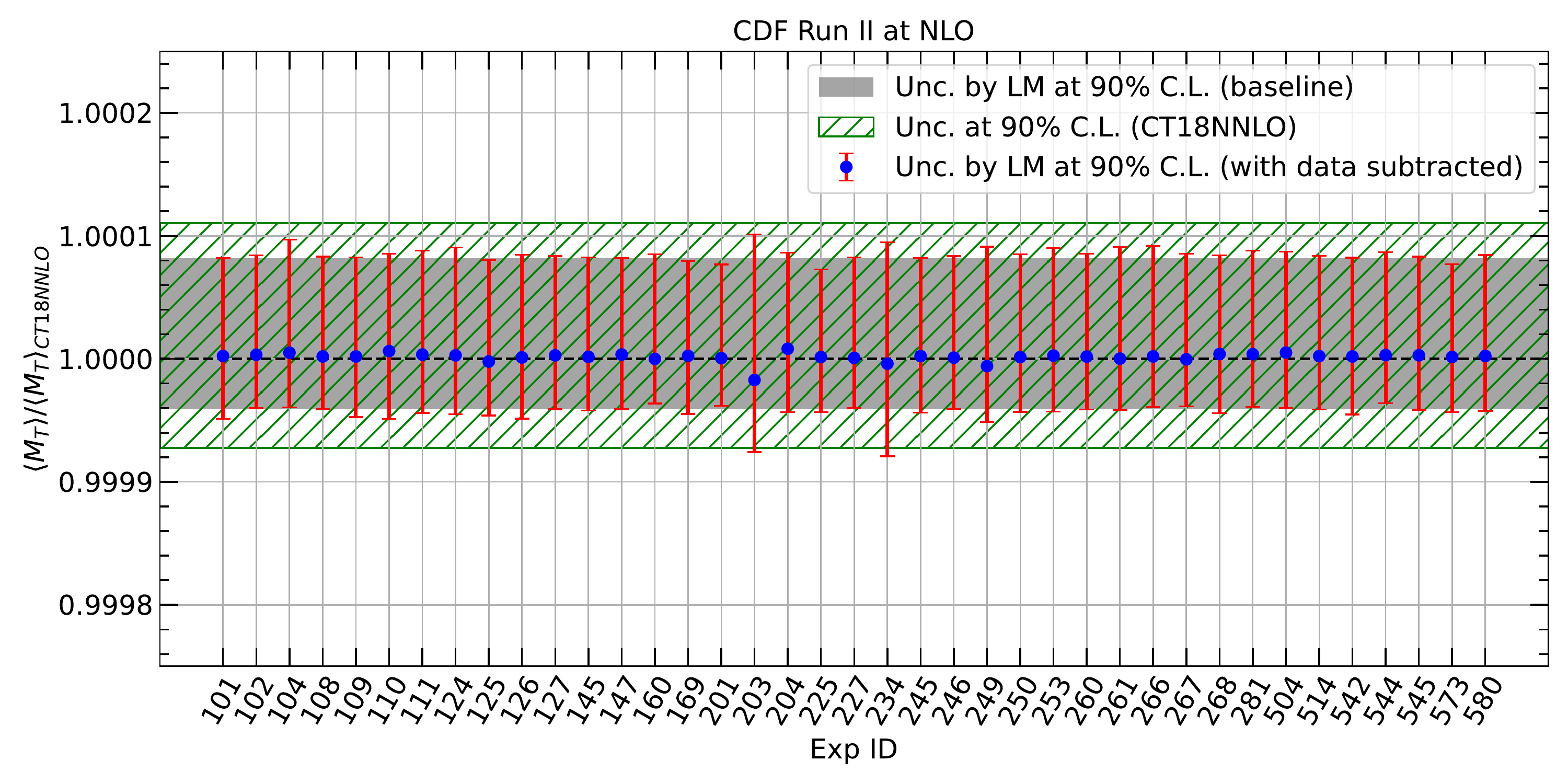}
		\includegraphics[width=0.7\textwidth]{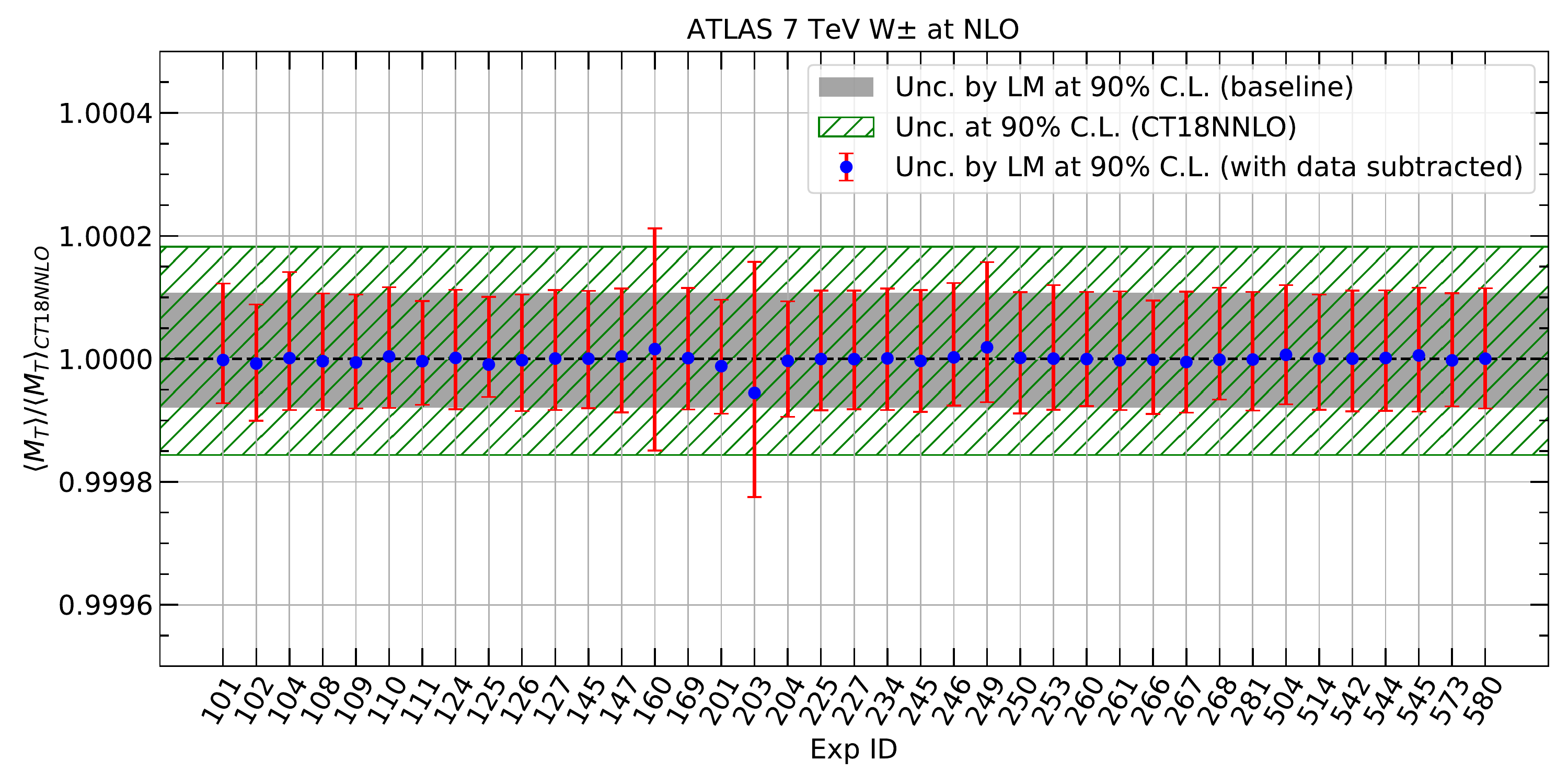}
		\caption{The results of LM scans on $\langle M_T\rangle$ with data subtracted.
			The results are normalized to the central value determined with full data sets.
			The horizontal axis represents the experimental data set removed from the LM scans.
			The blue mark and the red error bar respectively indicate the central value and uncertainties at 90\% C.L. determined with the LM method with the rest of the data sets.
			The green hatched area and the gray band represent the uncertainties at 90\% C.L. determined with the Hessian method and the LM method with full data sets respectively.}
		\label{fig:LM_sub}
	\end{figure}

	The contribution from an individual experimental data set can also be evaluated by LM scans
	with data subtracted.
	We remove one data set at a time and repeat the LM scans
	on $\langle M_T \rangle$ with the rest of the data sets.
	The difference between the fit with and without the data set can be an assessment of its contribution.
	In Fig.~\ref{fig:LM_sub}, we show the results of LM scans on the $\langle M_T\rangle$ with data subtracted.
	The results are normalized to the central value determined with full data sets.
	For the case of the CDF measurement, in the upper panel, we find that after the removal of most of the data set, the $\langle M_T\rangle$ value and its uncertainty are only changed slightly, as represented by those error bars compared to the uncertainty from LM scans with full data sets represented by the gray band.
	The E866 Drell-Yan ratio data (Exp. ID = 203) together with the D0 Run II charge asymmetry data (Exp. ID = 234) and the NMC deuteron data (Exp. ID = 104) give strong constraints, which is consistent with the left panel of Fig.~\ref{fig:LM_mt}.
	In addition, we find that the E866 Drell-Yan ratio data prefer a larger $\langle M_T\rangle$.
	For the case of the $W^{\pm}$ production at ATLAS 7 TeV, in the lower panel, the constraints from the HERA inclusive DIS data (Exp. ID = 160), the E866 Drell-Yan ratio data (Exp. ID = 203) and the CMS 8 TeV charge asymmetry data (Exp. ID = 249) predominate as expected.
	After the inclusion of the HERA inclusive DIS data or the E866 Drell-Yan ratio data, the uncertainties of $\langle M_T\rangle$ are reduced by almost 50\%.
	In addition, the E866 Drell-Yan ratio data prefer a larger $\langle M_T\rangle$ contrasted with the HERA inclusive DIS data and CMS 8 TeV charge asymmetry data which prefers a smaller value.

	\subsection{Discussions}~\label{sec:dis}

\begin{figure}
		\includegraphics[width=0.43\textwidth]{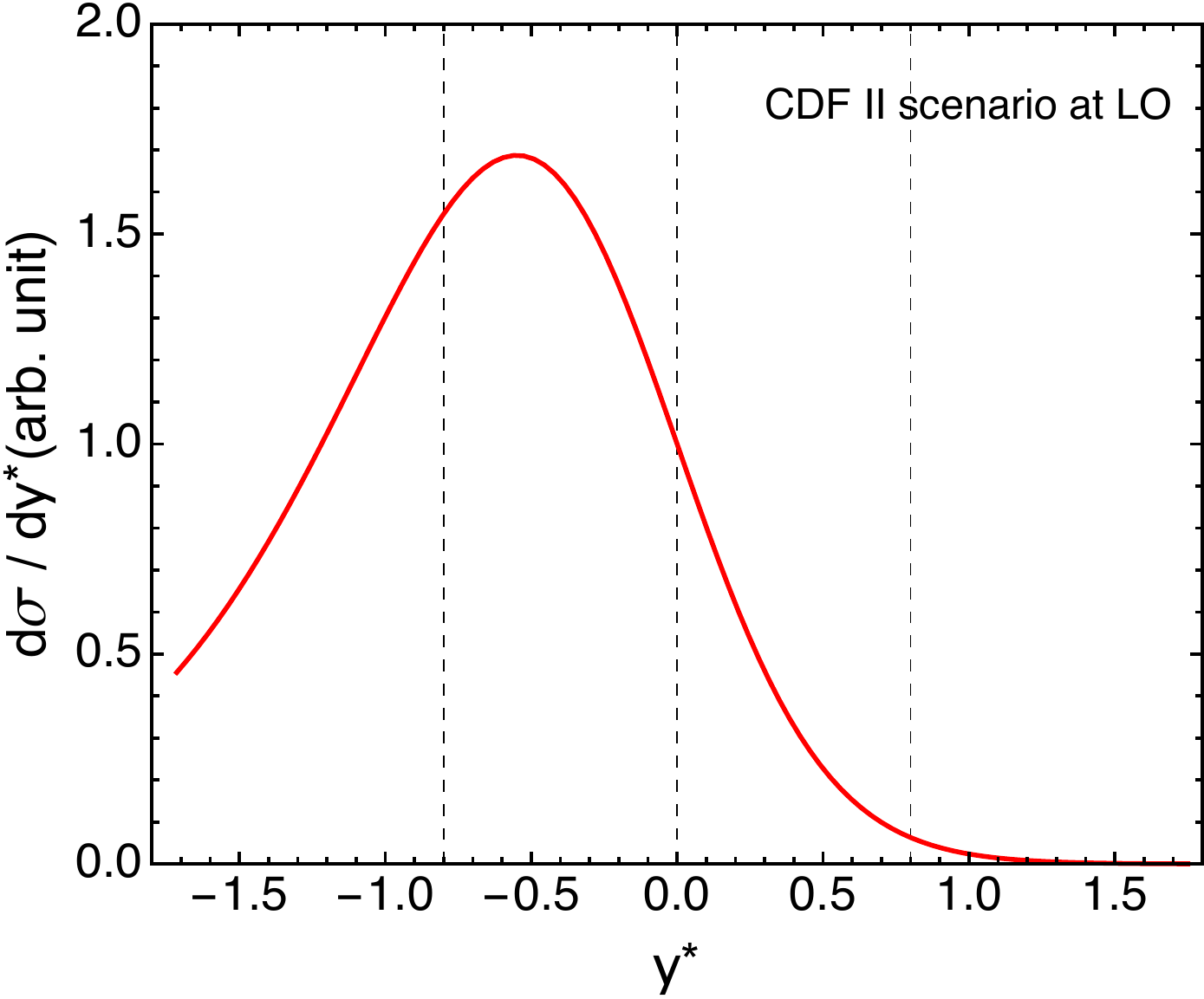}
		\hspace{0.3in}
		\includegraphics[width=0.5\textwidth]{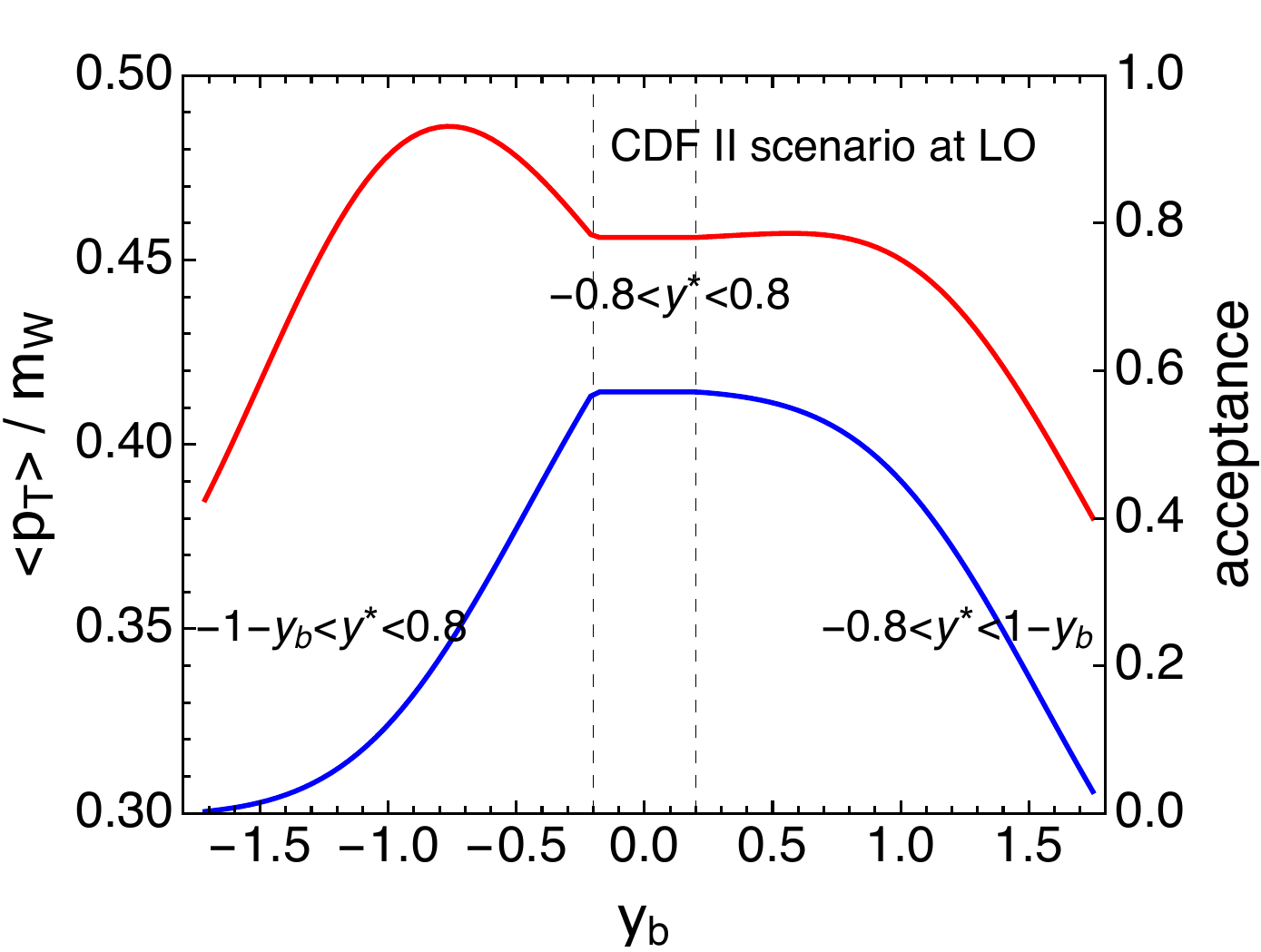}
		\caption{
			Left: rapidity distribution of the positron in the rest frame of the $W^+$ boson
			for the scenario of the CDF measurement calculated at LO for the $u\bar d$
			partonic channel;
			right: mean transverse momentum (red color with a scale on the left)
			and acceptance (blue color with a scale on the right) of the positron after
			selections as a function of the rapidity of the $W$ boson.
			The inequalities indicate the $y^*$ region that contribute to the
			cross sections after selections for a fixed $y_b$.}
		\label{fig:ptyb}
	\end{figure}

	In the following, we perform a simple analysis to further understand the dependences of the leptonic distributions on the PDFs focusing on $p_T$ of the charged lepton for
	the CDF scenario.
	Note at the LO $p_T$ equals half of the transverse mass discussed earlier.
	We defined $\theta^*$ as the polar angle between the decayed positron and the
	anti-proton directions, and $y^*$ as the rapidity of the positron with respect
	to the proton direction, both in the rest frame of the $W^+$ boson.
	The following relations on kinematics hold at the LO,
	\begin{equation}
		\cos \theta^*=-\tanh{y^*},\,\,p_T=\frac{m_W}{2}\frac{1}{\cosh{y^*}}.
	\end{equation}
	For the dominant partonic channel of $u\bar d$ annihilation, the LO partonic differential 
	cross section with respect to $\cos\theta^*$ is proportional to
	$(1+\cos\theta^*)^2$ due to the $V-A$ structure of the weak-charged current.
	That can be translated into
	\begin{equation}
		\frac{d\hat{\sigma}}{dy^*}\sim (1-\tanh y^*)^2/\cosh{^2y^*},
	\end{equation}
	and a $p_T$ weighted distribution
	\begin{equation}
		p_T\frac{d\hat{\sigma}}{dy^*}\sim \frac{m_W}{2} (1-\tanh y^*)^2/\cosh{^3y^*}.
	\end{equation}
	From the above equations, one can calculate the mean transverse momentum of the positron
	equals $\frac{15\pi}{128}\, m_W\sim 0.368\, m_W$.
	However, in the CDF analysis, due to the selection cut on the transverse momentum
	of the charged lepton, that imposes a cut of $|y^*|\lesssim 0.8$.
	Furthermore, if the $W$ boson is boosted with a rapidity of $y_b$, the rapidity
	selection of the charged lepton also requires $|y^*+y_b|<1$.
	Thus the actual mean transverse momentum and acceptance of the charged lepton
	depend on $y_b$ in a non-trivial way as plotted in Fig.~\ref{fig:ptyb}.
	We also show the distribution in $y^*$ in Fig.~\ref{fig:ptyb} which peaks at a
	negative value since the positron is aligned with the $\bar d$ quark.
	The inequalities indicate the $y^*$ region that contributes to the
	cross sections after selections for a fixed $y_b$.
	The mean transverse momentum peaks at a $y_b$ value of about $-0.8$ since
	that excludes the peak region in $y^*$ which has a relatively smaller $p_T$.
	The mean transverse momentum drops at large $|y_b|$ values because there
	only high $|y^*|$ regions are left which have rather low $p_T$.   
	The acceptance is largest in the central region of $y_b$.
	
	\begin{figure}
		\includegraphics[width=0.6\textwidth]{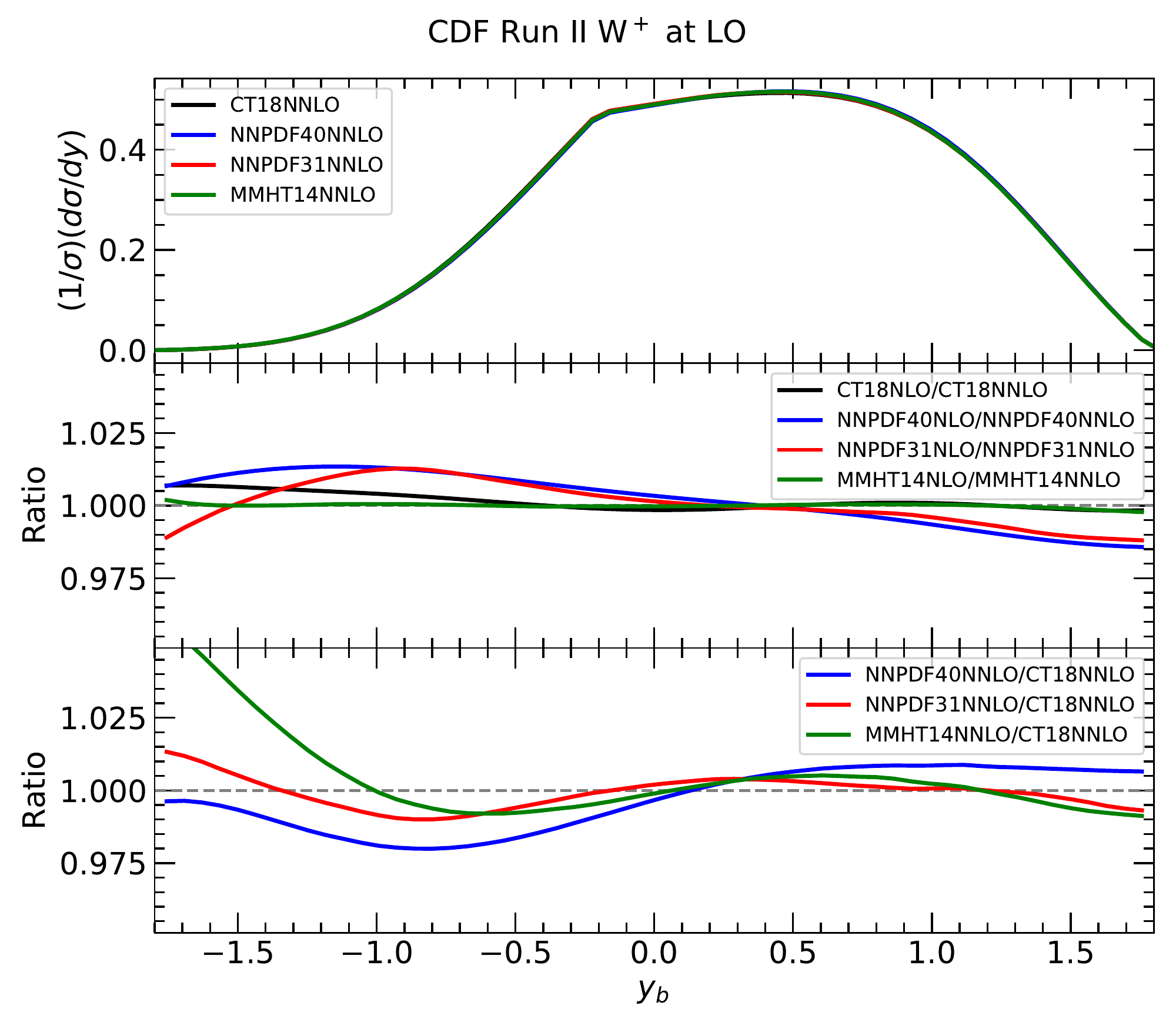}
		\caption{
			Rapidity distribution of the $W^+$ boson at Tevatron Run II calculated at the LO and with lepton acceptance applied, for $u\bar d$ annihilation channel using a variety of NLO and NNLO PDFs.
		}
		\label{fig:yw}
	\end{figure}

	On another hand, the transverse momentum distribution is a superposition of
	contributions from all possible rapidity of the $W$ boson.
	The distribution of the later is determined exactly by PDFs as
	\begin{equation}
		\frac{d{\sigma}}{dy_b}\sim f_{u/p}(x_1, \mu)f_{\bar d/\bar p}(x_2, \mu),
	\end{equation}
	at the LO for the $u\bar d$ partonic channel, with $x_{1,2}=m_W/\sqrt s\, e^{\pm y_b}$.
	The normalized rapidity distribution of the $W$ boson based on the above equation
	is shown in Fig.~\ref{fig:yw} for a variety of PDFs, with lepton acceptance
	shown in Fig.~\ref{fig:ptyb} applied.  
	By combining Figs.~\ref{fig:ptyb} and~\ref{fig:yw} we are now able to understand
	several features shown in Sec.~\ref{sec:cdf}. 
	For instance, by comparing the $y_b$ distributions from CT18, MSHT20, NNPDF3.1,
	and NNPDF4.0 NNLO PDFs, we can see the ratios of NNPDF4.0 to others show a clear positive
	slope across $y_b$.
	That leads to a reduction of the mean transverse momentum of the
	positron since it is in an average smaller for $y_b>0$, and also 
	a reduction of the mean transverse mass of the leptons.
	The behavior of NNPDF4.0 can be traced back as due to a suppression of the
	$d$-quark PDF at large-$x$ region compared to its previous generation,
	which suppresses the rapidity distribution in the anti-boost region.
	Similarly, it can be explained that the NLO PDFs give a larger mean transverse mass
	than NNLO PDFs in general, as can be seen, the ratios of NLO to NNLO PDFs show a negative slope from the middle panel of Fig.~\ref{fig:yw}.
	However, this situation can change for different PDF groups as CT18 and MMHT14
	show more stable predictions when using NLO PDFs compared to NNLO. 

	\section{Summary}~\label{sec:C}
	In summary, we study the dependence of the transverse mass distribution of the
	charged lepton and the missing energies on PDFs focusing on the $W$ boson
	mass measurements by the CDF and ATLAS collaborations.
	We compare the shape of the distribution for predictions using various up-to-date
	PDFs.
	In particular, we compare the mean transverse mass adapted to each of
	the measurement.
	We find that spread of predictions from different PDF sets can be much
	larger than the PDF uncertainty predicted by a specific PDF set.
	The mean transverse mass is strongly anti-correlated with the extracted
	$W$ boson mass as validated by the $\chi^2$ fit and by comparing
	to experimental numbers of mass shift on PDF dependence.
	Thus we suggest analyzing the experimental data using up-to-date PDFs could be
	highly desirable, especially considering tensions between different $W$ boson mass
	measurements.
	We also examine theoretical uncertainties induced by factorization and renormalization scales, the strong coupling constant, and the $W$-boson decay width in the CDF context and 
	find the width effect can be comparable to the experimental full uncertainty. 
	We further carry out a series of Lagrange multiplier scans to identify the constraints
	on the transverse mass distribution imposed by individual data sets in the CT18
	global analysis of PDFs.
	In the case of the CDF measurement, the distribution is mostly sensitive to the
	$d$-quark PDFs at intermediate $x$ region that are largely constrained by the DIS and
	Drell-Yan data on deuteron target, as well as the Tevatron lepton charge asymmetry data.  
	For the ATLAS measurement, the strongest constraints arise from the HERA inclusive DIS
	data, the E866 Drell-Yan ratio data, and the CMS lepton charge asymmetry data.

	\acknowledgments
	We would like to thank Pavel Nadolsky, C.-P. Yuan, as well as other CTEQ-TEA collaborators for helpful discussions and comments. JG would like to thank Hong-Jian He for useful discussions. 
	KX would give thanks to PITT PACC colleagues for many stimulating discussions, in particular to Tao Han for bringing up the $W$ decay width effect. 
	The work of JG and DL is supported by the National Natural Science Foundation of China under Grants No. 11875189 and No. 11835005. 
	The work of KX is supported by the U.S. Department of Energy under grant No. DE-SC0007914, U.S. National Science Foundation under Grant No. PHY-2112829, and in part by the PITT PACC.
	
	\bibliographystyle{utphys}
	\bibliography{CDF2MW}
\end{document}